\documentclass[aps,onecolumn,showpacs]{revtex4}
\usepackage{dcolumn}
\usepackage{graphicx}
\usepackage{amsmath}
\usepackage{amsfonts}
\usepackage{amssymb}
\usepackage{psfrag}
\usepackage{wrapfig}
\usepackage{subfigure}
\usepackage{makeidx}
\usepackage{bm}
\usepackage{epsf}

\newtheorem{lem}{Lemma}
\newtheorem{prop}{Proposition}

\begin{document}

\title{A focusing and defocusing semi-discrete complex short pulse equation and its varioius soliton solutions}
\author{Bao-Feng Feng$^{1}$}
\author{Liming Ling$^{2}$}
\author{Zuonong Zhu$^3$}
\affiliation{$^1$ School of Mathematical and Statistical Sciences, The University of
Texas Rio Grande Valley Edinburg TX, 78541-2999, USA}
\affiliation{$^2$ Department of Mathematics, South China University of Technology,
Guangzhou 510640, China}
\affiliation{$^3$ School of Mathematical Sciences, Shanghai Jiaotong University, Shanghai
200240, China}

\begin{abstract}
In this paper, we are concerned with a a semi-discrete
complex short pulse (CSP) equation of both focusing and defocusing types, which can be viewed as an analogue to the Ablowitz-Ladik (AL) lattice in the ultra-short pulse regime. By using a generalized Darboux transformation method, various solutions to this newly integrable semi-discrete equation are studied with both zero and nonzero boundary conditions. To be specific, for the focusing CSP equation, the multi-bright solution (zero boundary condition), multi-breather and high-order rogue wave solutions (nonzero boudanry conditions) are derived, while for the defocusing CSP equation with nonzero boundary condition, the multi-dark soliton solution is constructed. We further show that, in the continuous limit, all the solutions obtained converge to the ones for its original CSP equation (see Physica D, 327 13-29 and Phys. Rev. E 93 052227)
%As the application, the multi-soliton, multi-breather and high-order rogue wave solutions are derived through the generalized Darboux transformation.
\newline
\textbf{Key words:} generalized Darboux transformation, semi-discrete complex short
pulse equation, bright and dark soliton, breather, rogue wave \newline
\textbf{Mathematics Subject Classification:} 39A10, 35Q58
\end{abstract}
\maketitle

%%%%%%%%%%%%%%%%%%%%%%%%%%%%%%%%%%%%%%%%%%%%%%%%%

%\date{Sept 22, 2015}
%\date{Dec 28, 2018}

\section{Introduction}
The most recent advances in nonlinear optics include the generation and applications of ultrashort optical pulses, whose time duration is typically of the order of femtoseconds, which lead to the Nobel prize in physics 2018 \cite{2018Nobelprize}. It has been an important topics for the mathematical models of optical pulse propagation in medium \cite{Agrawalbook,HasegawaKodama,AgrawalKivsharbook}. When the range of short pulse width is from 10 $ns$ to 10 $fs$, both dispersive
and nonlinear effects influence their shape and spectrum. The following full wave equation
\begin{equation}
\nabla ^{2}\mathbf{E}-\frac{1}{c^{2}}\mathbf{E}_{tt}=\mu _{0}\mathbf{P}%
_{tt}\,,  \label{E-wave-equation}
\end{equation}%
originates directly from Maxwell's equation. If we assume the local medium
response and only the third-order nonlinear effects (Kerr effect)
%governed by $\chi ^{(3)}$,
the induced polarization consists of linear and nonlinear parts, $\mathbf{P}(\mathbf{r},t)=\mathbf{P}_{L}(\mathbf{r},t)+\mathbf{P}_{NL}(\mathbf{r},t)$.
%where the linear part
%\begin{equation}
%\mathbf{P}_{L}(\mathbf{r},t)=\epsilon _{0}\int_{-\infty }^{\infty }\chi
%^{(1)}(t-t^{\prime })\cdot \mathbf{E}(\mathbf{r},t^{\prime })\,dt^{\prime
%}\,,  \label{P_L}
%\end{equation}%
%and the nonlinear part
%\begin{equation}
%\mathbf{P}_{NL}(\mathbf{r},t)=\epsilon _{0}\int_{-\infty }^{\infty }\chi
%^{(3)}(t-t_{1},t-t_{2},t-t_{3})\times %\mathbf{E}(\mathbf{r},t_{1})\mathbf{E}(%
%\mathbf{r},t_{2})\mathbf{E}(\mathbf{r},t_{3})\,dt_{1}dt_{2}dt_{3}\,.
%\label{P_NL}
%\end{equation}

Under the assumption of quasi-monochromatic, i.e., the pulse spectrum, centered as $\omega_{0}$, is assumed to have a spectral width $\Delta \omega $ such that $\Delta \omega /\omega _{0}<<1$. Under this assumption, one can expand the
frequency-dependent dielectric constant $\epsilon (\omega )$ in terms of
Taylor series up to the second order. As a result, the so-called nonlinear Schr\"odinger equation
\begin{equation}
\mathrm{i}q_{z}+\alpha _{1}q_{\tau \tau }+\alpha _{2}|q|^{2}q=0\,,
\label{NLS}
\end{equation}
can be derived to govern the slowly varying envelop of optical waves in weakly nonlinear dispersive media \cite%
{HasegawaTappert1, HasegawaTappert2}.  Here $\alpha _{1}$ represents the effect of group velocity
dispersion (GVD) ($\alpha _{1}>0$ corresponds to anomalous dispersion, also
focusing case, and $\alpha _{1}<0$--normal dispersion, also defocusing case), $\alpha _{2}$ represents the self-phase modulation (SPM) due to Kerr effect.

Upon switching the spatial and temporal variables and normalization, the nonlinear Schr\"{o}dinger (NLS) equation can be put into a standard form
 \begin{equation}  \label{NLS1}
{\mathrm i} q_{t}=q_{xx}+2 \sigma |q|^{2} q\,
\end{equation}
which has become a generic model equation describing the evolution of small amplitude and slowly varying wave packets in weakly nonlinear media \cite{Agrawalbook,HasegawaKodama,AgrawalKivsharbook,Ablowitzbook,APT}. It arises in a variety of physical contexts such as nonlinear optics metioned above, Bose-Einstein condesates \cite%
{BECReview}, water waves \cite{Benney1967} and plasma physics \cite%
{ZakharovPlasma}. The integrability, as well as the bright-soliton solution in the focusing case ($\sigma=1$), was found by Zakharov and Shabat \cite{ZakharovShabat,ZakharovShabat2}. The dark soliton was found in the defocusing NLS equation  ($\sigma=-1$) \cite{HasegawaTappert2}, and was observed experimentally in 1988 \cite{Krokeldark1,Weinerdark1}.  Recently, the rogue (freak) waves were discovered and the
Peregrine soliton was found in focusing NLS equation \cite%
{Opticalroguewave,Peregrinesoliton}.

The integrable discretization of nonlinear Schr\"{o}dinger equation
\begin{equation} \label{AL}
\mathrm{i}q_{n,t}=\left( 1+\sigma |q_{n}|^{2}\right) \left(
q_{n+1}+q_{n-1}\right) \,
\end{equation}
was originally derived by Ablowitz and Ladik \cite{AblowitzLadik,AL2}, so it is also called the Ablowitz-Ladik (AL) lattice equation. Similar to the continuous case, it is known that the AL lattice equation, by Hirota's bilinear method, admits the bright soliton solution for the focusing case ($\sigma =1$) \cite{TsujimotoBookchapter,Narita1990}, also the dark soliton solution for the defocusing case ($\sigma =-1)$ \cite{OhtaMaruno}. The inverse scattering transform (IST) has been developed by several authors in the literature \cite{Konotop92,ABB07,Mee15,BPrinari,BPrinari2}. The rogue wave solutions to the AL lattice and coupled AL lattice equations were constructed by Darbourx transformation method \cite{Akhmediev1,Akhmediev2} and by Hirota's bilinear method \cite{OhtaYangAL}.
The geometric construction of the AL lattice equation was given by Doliwa and Santini \cite{DoliwaAL}.

When the width of optical pulses is less than 1 $ps$, higher order
nonlinear effects has to be taken into account, and the NLS equation should
be modified. As a result, a generalized NLS (gNLS) equation \cite{HasegawaKodama}
\begin{equation}
\mathrm{i}q_{z}+\alpha _{1}q_{\tau \tau }+\alpha _{2}|q|^{2}q+\mathrm{i}%
(\beta _{1}q_{\tau \tau \tau }+\beta _{2}|q|^{2}q_{\tau }+\beta
_{3}q(|q|^{2})_{\tau })=0\,,
\end{equation}%
can be derived, where $\beta _{1}$, $\beta _{2}$ and $\beta _{3}$ are the
parameters related to the third order dispersion (TOD), self-steepening (SS) and stimulated Raman scattering (SRS). Due to the complexity of the gNLS equation, the study is mainly numerical. However, in some special case, the gNLS equation becomes integrable and is available for rigorous analysis. For example, when $\beta _{1}=\beta _{3}=0,$the gNLS equation is called modified NLS equation, which is integrable.
\begin{equation}
\mathrm{i}q_{z}+\alpha _{1}q_{\tau \tau }+\alpha _{2}|q|^{2}q+\mathrm{i}%
\beta _{2}|q|^{2}q_{\tau }=0\,.
\end{equation}%
In addition, there are two known integrable cases when $\beta _{1}\neq
0,\beta _{3}\neq 0,$ and the condition $3\beta _{1}\alpha _{2}=\beta
_{2}\alpha _{1}$ is satisfied. Under this condition, a gauge transformation
brings the generalized NLS equation into the form
\begin{equation}
q_{z}+\beta _{1}^{^{\prime }}q_{\tau \tau \tau }+\beta _{2}^{^{\prime
}}|q|^{2}q_{\tau }+\beta _{3}^{^{\prime }}q(|q|^{2})_{\tau }=0\,.
\end{equation}%
When $\beta _{1}^{{\prime}}:\beta _{2}^{{\prime }}:\beta_{3}^{{\prime}}=1:6:0$, we obtain Hirota's equation \cite{hirotaeq}. When $\beta_{1}^{{\prime }}:\beta _{2}^{{\prime }}:\beta_{3}^{{\prime }}=1:6:3$, we find the Sasa-Satsuma equation \cite{sasasatsuma}.

However, when the width of optical pulse is of the order
of sub-femtosecond ($<10^{-15}$ s), then the width of its spectrum is of the order greater than $10^{15}$ $s^{-1}$, being comparable with the spectral width of the optical pulses,
the quasi-monochromatic assumption is not valid
anymore.
%We need to construct a suitable fit to $\epsilon (\omega )$ in the
%desired spectral range. %\include{00-summaryFeng}
Instead, by using the Kramers-Kronig relation of the response function
%can be expressed into
%\begin{equation}
% \epsilon(\omega) =\bar\epsilon\left(1- \frac{\omega^2_0}{\omega^2} %\right)\,,
%\end{equation}
one obtains a normalized model equation
\begin{equation}
E_{zz}-E_{tt}= E+ \left( |E|^{2}E\right) _{tt}\,,  \label{Wave_nonlinear1b}
\end{equation}%
and further a non-integrable equation
 \begin{equation}
2E_{z\tau}= E+ \left( |E|^{2}E\right) _{\tau\tau}\,,  \label{Wave_nonlinear2b}
\end{equation}%
where $\tau=t-z$. The solitary wave solutions and their interactions for above models were studied in details in
\cite{Rothenberg,Skobelev,Kim,Amir,Amir2}.

Following the same spirit, Sch\"{a}fer and Wayne proposed a so-called
short-pulse (SP) equation \cite{SPE_Org} %\begin{equation}
%u_{xt}=u+ \frac 16 \left(u^{3}\right)_{xx}\,,  \label{SPE}
%\end{equation}%
to describe the propagation of ultrashort optical pulses in silicon fiber.
\begin{equation}
u_{xt}=u+\frac{1}{6}\left( u^{3}\right) _{xx}\,.  \label{SPE}
\end{equation}
Here, $u=u(x,t)$ is a real-valued function, representing the magnitude of
the optical field.
%It has attracted much attention ever since the SP equation was proposed.
It is completely integrable \cite{Sakovich}, admits periodic and soliton
solutions \cite{Matsuno_SPEreview}. Wave breaking phenomenon was studied in \cite{SPWB}. The integrable discretization of the SP
equation and its geometric formulation was studied in \cite%
{d-short,SPE_discrete2}. %%

However, similar to the NLS equation, the complex representation has
advantages for the description of the optical waves since a single
complex-valued function can contains the information of amplitude and phase
in a wave packet simultaneously. Consequently, a complex short pulse (CSP) equation
%\cite {,LingFeng2}
\begin{equation}
q_{xt}+q+\frac{1}{2}\sigma \left( |q|^{2}q_{x}\right) _{x}=0\,  \label{CSP}
\end{equation}%
was proposed by the authors \cite{Feng_ComplexSPE,FenglingzhuPRE}.
Here $q=q(x,t)$ is a complex-valued function, representing the optical wave packets in ultrashort pulse regime, $\sigma =\pm 1$ where $+1$ represents the focusing case, and $-1$ stands for the defocusing case.  It can be viewed as an analogue of the nonlinear Schr\"{o}dinger equation in ultra-short pulse regime.

%As is shown in \cite%
%{Feng_ComplexSPE}, the focusing CSP equation admits the pfaffian-type
%multi-soliton solutions.

For the focusing CSP equation, its multi-bright solition solution has been
found in pfaffian form in \cite{Feng_ComplexSPE} and in determinant form in \cite%
{FengShen_ComplexSPE} by combing Hirota's bilinear method and the
Kadomtsev--Petviashvili (KP) hierarchy reduction method. In addition to
above multi-bright soliton solution, the multi-breather and the higher order
rogue wave solutions are constructed via Darboux tranformation method \cite%
{LFZPhysD}. For the defocusing CSP equation, its multi-dark soliton solution
is constructed by the KP hierarchy reduction method \cite{FMO_ComplexSPE}
and generalized Darboux transformation method \cite{FenglingzhuPRE},
respectively. Periodic and soliton solutions for both the focusing and defocusing CSP equation are studied in \cite{Optik}.  The long time asymptotics were analyzed for both the real and complex short pulse equation in \cite{Jxu1,Jxu2}.
In \cite{FengShen_ComplexSPE,FMO_ComplexSPE}, the geometric
formulation of the CCD equation and a geometric interpretation for the
hodograph transformation was given for the focusing and defocusing CSP
equation, respectively. It is noted that if we exchange the coordinates $x$
and $t$, the CCD system becomes the complex sine-Gordon equation \cite{NolsG}
or the AB system \cite{AB}, which is the first negative flow of the AKNS
system.

Recently, much attention has been paid to the study of discrete integrable systems \citep{Disbook}
%J. Hietarinta, N. Joshi, F.W. Nihoff \textit{Discrete Systems and Integrability},
%(Cambridge University Press, 2016).
though it can traced back to the middle of 1970s when Hirota discretized various famous soliton equations such as the KdV, mKdV, and the sine-Gordon equations through the bilinear method \cite{hirota}. In the past two decades, the field of discrete system has grown to prominence as an area in which numerous breakthroughs have taken place, inspriring new developments in other areas of mathematics.
As mentioned previously, Ablowitz and Ladik proposed a method of integrable
discretizations soliton equations which involve the nonlinear Schr\"{o}%
dinger equations and mKdV equation, by the Lax pair method \cite{Ablowitz}.
Another successful way to discretize soliton equations was proposed by Date
et al via the transformation group theory, which gives a large number of
integrable disretizations \cite{date}. One of the most interesting examples
is the discrete KP equation, or the so-called Hirota-Miwa equation, which embodies the whole KP hierarchy \cite{Hirota:dKP,Miwa}.
Suris also developed a general Hamiltonian approach for integrable discretizations of integrable systems \cite{suris}. Following the original works on the integrable discretizations, the study on discrete integrable systems has been extended to other mathematics fields such as discrete differential
geometry \cite{bobenko}.

In the present paper, we are concerned with various soliton solutions to a
semi-discrete complex short pulse equation
\begin{equation}
\frac{d}{dt}\frac{q_{n+1}-q_{n}}{\Delta x_{n}}+\frac{1}{2}(q_{n+1}+q_{n})+%
\frac{\sigma }{2}\frac{1}{\Delta x_{n}}\left( |q_{n+1}|^{2}\frac{%
q_{n+2}-q_{n+1}}{\Delta x_{n+1}}-|q_{n}|^{2}\frac{q_{n+1}-q_{n}}{\Delta x_{n}%
}\right) =0.  \label{sem-csp}
\end{equation}
where $q_{n}=q(nx_{n},t),\Delta x_{n}=x_{n+1}-x_{n}$. This lattice equation is a semi-discrete analogue of the complex short pulse (CSP) equation, where the spatial variable is discretized and the time variable remains continuous.
%\begin{equation}
%q_{XT}+q+\frac{1}{2}\sigma (|q|^{2}q_{X})_{X}=0,\quad \sigma =\pm 1.
%\label{csp}
%\end{equation}%
%The CSP equation
The semi-discrete CSP equation can also be written in a coupled
two-component system \cite{FMO-PJMI,FMOmultiSP}
\begin{equation} \label{sem-csp2}
\begin{split}
& \frac{d}{dt}(q_{n+1}-q_{n})=\frac{1}{2}(x_{n+1}-x_{n})(q_{n+1}+q_{n}), \\
& \frac{d}{dt}(x_{n+1}-x_{n})+\frac{1}{2}\sigma
(|q_{n+1}|^{2}-|q_{n}|^{2})=0\,.
\end{split}%
\end{equation}

%The semi-discrete CSP equation is integrable becasue it has a Lax
%pair in the form
%\begin{equation}
%\begin{split}
%\Psi _{n+1}=& U_{n}\Psi _{n}, \\
%\Psi _{n,t}=& V_{n}\Psi _{n},
%\end{split}%
%\end{equation}%
%where
%\begin{equation*}
%U_{n}=%
%\begin{bmatrix}
%1-\frac{\mathrm{i}(X_{n+1}-X_{n})}{\lambda } & -\sigma \frac{q_{n+1}^{\ast
%}-q_{n}^{\ast }}{\lambda } \\[8pt]
%\frac{q_{n+1}-q_{n}}{\lambda } & %1+\frac{\mathrm{i}(X_{n+1}-X_{n})}{\lambda }
%\\
%&
%\end{bmatrix}%
%,\,\,V_{n}=\frac{\mathrm{i}}{4}\lambda \sigma _{3}+\frac{\mathrm{i}}{2}%
%Q,\,\,Q=%
%\begin{bmatrix}
%0 & \sigma q_{n}^{\ast } \\[8pt]
%q_{n} & 0 \\
%&
%\end{bmatrix}%
%\,.
%\end{equation*}
%The compatability condition gives the semi-discrete CSP equation.
As mentioned in \cite{FMO-PJMI}, the semi-discrete CSP equation can be constructed in a very direct way. It is shown in \cite{FenglingzhuPRE,FMO_ComplexSPE} that the CSP equation is
related to the so-called complex coupled dispersionless (CCD) equation
\begin{equation}
%\begin{split}
 q_{ys}=q\rho, \quad  \rho_{s}+\frac{1}{2}\sigma (|q|^{2})_{y}=0,
%\end{split}
\label{cd-c}
\end{equation}%
by a reciprocal (hodograph) transformation defined by $\mathrm{d}x=\rho
\mathrm{d}y-\frac{1}{2}\sigma |q|^{2}\mathrm{d}s$, $\mathrm{d}t=-\mathrm{d}s$. The CCD equation \cite{LFZPhysD,FenglingzhuPRE} admits a Lax pair of the
form
\begin{equation}
\Psi _{y}=U(\rho ,q;\lambda )\Psi, \quad  \Psi _{s}=V(q;\lambda )\Psi, \label{ccd-lax}
\end{equation}%
%\begin{equation}
%  \label{ccd-laxb}
%\end{equation}%
where
\begin{equation}
U(q,\rho ;\lambda )=\lambda ^{-1}%
\begin{bmatrix}
-\mathrm{i}\rho & -\sigma {q^{\ast }}_{y} \\[8pt]
q_{y} & \mathrm{i}\rho%
\end{bmatrix}%
,\quad \,V(q;\lambda )=\frac{\mathrm{i}}{4}\lambda \sigma _{3}+\frac{\mathrm{%
i}}{2}Q,
\end{equation}%
with $^{\ast }$ representing the complex conjugate, $\sigma _{3}$, the
third Pauli matrix, and $Q$ being
\begin{equation*}
\sigma _{3}=%
\begin{bmatrix}
1 & 0 \\[8pt]
0 & -1%
\end{bmatrix}%
,\quad \,Q=%
\begin{bmatrix}
0 & -\sigma {q^{\ast }} \\[8pt]
q & 0%
\end{bmatrix}%
\,,
\end{equation*}%
respectively. Replacing the forward-difference to the first-order derivative in the spatial part of the Lax pair, i.e.,
\begin{equation*}
\frac{\Psi _{n+1}-\Psi _{n}}{a}={\lambda }^{-1}%
\begin{bmatrix}
-\mathrm{i}\rho _{n} & -\sigma \frac{q_{n+1}^{\ast }-q_{n}^{\ast }}{a} \\%
[8pt]
\frac{q_{n+1}-q_{n}}{a} & \mathrm{i}\rho _{n}
\end{bmatrix}%
\Psi _{n}\,,
\end{equation*}%
one yields the Lax pair for the semi-discrete CCD equation
\begin{equation}
\begin{split}
\Psi _{n+1}=& U_{n}\Psi _{n}, \\
\Psi _{n,s}=& V_{n}\Psi _{n},
\end{split}
\label{cd-lax}
\end{equation}%
where
\begin{equation}
U_{n}=%
\begin{bmatrix}
1-\frac{\mathrm{i}a\rho _{n}}{\lambda } & -\sigma \frac{q_{n+1}^{\ast
}-q_{n}^{\ast }}{\lambda } \\[8pt]
\frac{q_{n+1}-q_{n}}{\lambda } & 1+\frac{\mathrm{i}a\rho _{n}}{\lambda }
\end{bmatrix}%
,\,\,V_{n}=\frac{\mathrm{i}}{4}\lambda \sigma _{3}+\frac{\mathrm{i}}{2}%
Q_n,\,\,Q_n=%
\begin{bmatrix}
0 & \sigma q_{n}^{\ast } \\[8pt]
q_{n} & 0
\end{bmatrix}%
\,.
\end{equation}%
The compatibility condition gives exactly the semi-discrete CCD equation.
%\begin{equation}
%\begin{split}
%& \frac{q_{n+1,t}-q_{n,t}}{a}=\frac{1}{2}\rho _{n}(q_{n+1}+q_{n}), \\
%& \rho _{n,t}+\frac{\sigma }{2a}(|q_{n+1}|^{2}-|q_{n}|^{2})=0\,.
%\end{split}
%\label{cd}
%\end{equation}%
Replacing $a\rho _{n}$ by $x_{n+1}-x_{n}$, one obtains the semi-discrete CSP equation. As an analogue to the AL lattice equation in the ultra-short pulse regime, it is imperative to study this new integrable semi-discrete CSP equation due to its potential applications in physics. However, compared with the results for AL lattice equation, the obtained results for the semi-discrete CSP equation is much less. This motivates the present work, which intends to construct various soliton solutions with vanishing and non-vanishing boundary conditions via generalized Darboux transformation method.
%we need to discretize the hodograph transformation. The second
%equation in \eqref{cd} ensures the existence of $X_{n}(t)$. Moreover, it %can
%be solved exactly with the following form
%\begin{equation}
%X_{n}(t)=\rho _{-\infty }(-\infty )n-\frac{\sigma }{2}|q_{-\infty
%}(t)|^{2}t+\sum_{j=-\infty }^{n-1}a[\rho _{j}(-\infty )-\rho _{-\infty
%}(-\infty )]-\frac{\sigma }{2}\int_{-\infty %}^{t}[|q_{n}(s)|^{2}-|q_{-\infty
%}(s)|^{2}]\mathrm{d}s.  \label{Xn}
%\end{equation}

Darboux transformation, originating from the work of
Darboux in 1882 on the Sturm-Liouville equation, is a
powerful method for constructing solutions for integrable
systems \cite{Matveev}. However, the classical Darboux transformation cannot be iterated at the
same spectral parameter to obtain the multi-dark, breather and higher order rogue wave solutions. To overcome this difficulty, one of the authors  generalized the classical Darboux transformation by using a limit technique\cite{Guo1,Guo2},
which can be used to yield these solutions. It is noted that recently various soliton solutions have been found for the nonlocal NLS equation \cite{Taoxu1,Taoxu2}.
In this paper, we aim at finding soliton solutions for the semi-discrete CSP equation \eqref{sem-csp} by generalized Darboux transformation (DT).  It should be pointed out that the DT and soliton solutions for
the semi-discrete coupled dispersionless equation has been constructed by
Riaz and Hassen \cite{hassan} very recently. By the study for this system,
we find the discretization equation keeps Darboux transformation and
solitonic solutions with the original equation. Meanwhile, these results
would be useful to study the self-adaptive moving mesh schemes for the
complex pulse type equations.

The outline of the present paper is organized as follows. In section \ref%
{section2}, a generalized Darboux transformation of
semi-discrete CSP equation was derived through loop group method \cite%
{loop-group}. Based on the generalized Darboux transformation, we can obtain
the general solitonic formula for semi-discrete CSP equation. Moreover,
together with the reciprocal transformation, we can construct the general
solitonic formula for semi-discrete CSP equation in terms of the determinant
representation. The $N$-bright soliton solution for the focusing case with zero boundary condition and $N$-dark soliton solution for the defocusing case with nonzero boundary condition are constructed in section \ref{section3} and \ref{section4}, respectively. In \ref{section5}, the multi-breather solution with nonzero boundary condition is constructed for the focusing semi-discrete CSP and is approved to converge to the one for the original CSP equation in the continuous limit. Based on the multi-breather solution, we further derive general rogue wave solution in \ref{section6}.  Section \ref{section7} is devoted to conclusions and discussions.

\section{Generalized Darboux transformation for the semi-discrete CSP equation}
\label{section2}
%Semi-discrete coupled dispersion equation \eqref{cd} admits the following Lax pair
%\begin{equation}\label{cd-lax}
%    \begin{split}
%        \frac{\Psi_{n+1}-\Psi_n}{a}=& \Psi_n,  \\
%        \Psi_{n,t}=&V_n\Psi_n,
%    \end{split}
%\end{equation}
%where
%\begin{equation*}
%    U_n=\begin{bmatrix}
%                         1-\frac{{\rm i}a\rho_n}{\lambda} & -\sigma\frac{q^*_{n+1}-q^*_{n}}{\lambda} \\[8pt]
%                         \frac{q_{n+1}-q_{n}}{\lambda} & 1+\frac{{\rm i}a\rho_n}{\lambda} \\
%                      \end{bmatrix},\,\, V_n=\frac{{\rm i}}{4}\lambda\sigma_3+\frac{{\rm i}}{2}Q,\,\, Q=\begin{bmatrix}
%                   0  & \sigma q_n^* \\[8pt]
%                   q_n & 0 \\
%                 \end{bmatrix},
%\end{equation*}
%and $^*$ represents the complex conjugation.
Based on the Lax pair of the semi-discrete CSP equation \eqref{cd-lax},
%To obtain the solitonic formula for semi-discrete CD equation \eqref{cd},
we give the Darboux transformation by the following proposition.
%This proof
%of the proposition is the analogue in our previous work for the CSP %equation
%\cite{LFZPhysD}. Here we rewrite it to facilitate reading of some readers.
\begin{prop}
The Darboux matrix
\begin{equation}  \label{DT}
T_n=I+\frac{\lambda_1^*-\lambda_1}{\lambda-\lambda_1^*}P_n,\,\, P_n=\frac{%
|y_{1,n}\rangle \langle y_{1,n}|J}{\langle y_{1,n}|J|y_{1,n}\rangle},\,\, J=%
\mathrm{diag}(1,\sigma),
\end{equation}
can convert system \eqref{cd-lax} into a new system
\begin{equation*}
\begin{split}
\Psi_{n+1}^{[1]}=&U_n(\rho_n^{[1]},q_n^{[1]};\lambda)\Psi_{n}^{[1]}, \\
\Psi_{n,s}^{[1]}=&V_n(\rho_n^{[1]},q_n^{[1]};\lambda)\Psi_{n}^{[1]},
\end{split}%
\end{equation*}
where $|y_{1,n}\rangle=(\psi_{1,n},\phi_{1,n})^T$ is a special solution for
system \eqref{cd-lax} with $\lambda=\lambda_1$, $|y_{1,n}\rangle^{\dag}=%
\langle y_{1,n}|$. The B\"acklund transformations between $%
(\rho_n^{[1]},q_n^{[1]})$ and $(\rho_n,q_n)$ are given through
\begin{equation}  \label{backlund}
\begin{split}
\rho_n^{[1]}=&\rho_n-\frac{2}{a}\ln_{s}\left(\frac{E(\langle
y_{1,n}|J|y_{1,n}\rangle)}{\langle y_{1,n}|J|y_{1,n}\rangle}\right), \\
q_n^{[1]}=&q_n+\frac{(\lambda_1^*-\lambda_1)\psi_{1,n}^*\phi_{1,n}}{\langle
y_{1,n}|J|y_{1,n}\rangle}, \\
|q_n^{[1]}|^2=&|q_n|^2+4\sigma\ln_{ss}\left(\frac{\langle
y_{1,n}|J|y_{1,n}\rangle}{\lambda_1^*-\lambda_1}\right),
\end{split}%
\end{equation}
and the symbol $E$ denotes the shift operator $n\rightarrow n+1.$
\end{prop}

\textbf{Proof:} Firstly, we see that the evolution part of system %
\eqref{cd-lax} is a standard one for the AKNS system with $SU(2)$ symmetry.
Then the Darboux transformation for the AKNS system also satisfies the
system \eqref{cd-lax}. The rest of the proposition is to prove the formulas %
\eqref{backlund}. To derive these formulas, we borrow some idea from the
classical monograph \cite{alg-geo}.

Suppose there is a holomorphic solution for Lax pair equation \eqref{cd-lax}
in some punctured neighborhood of infinity on the Riemann surface, soothing
depending on $n$ and $s$, with the following asymptotical expansion at
infinity:
\begin{equation}  \label{infinity-asy}
\begin{bmatrix}
\psi_1 \\
\phi_1 \\
\end{bmatrix}%
=\left(%
\begin{bmatrix}
1 \\
0 \\
\end{bmatrix}%
+\sum_{i=1}^{\infty}\Psi_i\lambda^{-i}\right)\exp{\left(\frac{\mathrm{i}}{4}\lambda s\right)},\,\, \lambda\rightarrow \infty^{+}.
\end{equation}
On the one hand, from the Lax pair equation \eqref{cd-lax}, we have
\begin{equation}  \label{psiphi}
\begin{split}
\psi_{1,s}=&\frac{\mathrm{i}}{4}\lambda \psi_1+\frac{\mathrm{i}}{2}\sigma
q_n^*\phi_1, \\
\phi_{1,s}=&\frac{\mathrm{i}}{2}q_n\psi_1-\frac{\mathrm{i}}{4}\lambda \phi_1.
\end{split}%
\end{equation}
With the aid of above equations \eqref{psiphi}, we can obtain the following
relation
\begin{equation}
H_s=\frac{\mathrm{i}}{2}|q_n|^2-\frac{\mathrm{i}}{2}\lambda H-\frac{\mathrm{i%
}}{2}\sigma H^2+\ln_t(q_n^*)H,\,\, H\equiv q_n^*\frac{\phi_1}{\psi_1}=
\sum_{i=1}^{\infty}H_i\lambda^{-i}.
\end{equation}
Then the coefficient $H_i$ can be determined as following:
\begin{equation*}
\begin{split}
H_1=&|q_n|^2,\,\, H_2=2\mathrm{i}q_{n,s}q_n^*, \\
H_{i+1}=&2\mathrm{i}q_n^*\left(\frac{H_i}{q_n^*}\right)_s-\sigma%
\sum_{j=1}^{i-1}H_{j}H_{i-j},\,\, i\geq 2.
\end{split}%
\end{equation*}
It follows that the first equation of \eqref{psiphi} can be rewritten as
\begin{equation}
\psi_{1,s}=\left(\frac{\mathrm{i}}{4}\lambda +\frac{\mathrm{i}}{2}%
\sum_{i=1}^{\infty}H_i\lambda^{-i}\right)\psi_1.
\end{equation}
On the other hand, substituting the asymptotic expansion %
\eqref{infinity-asy}:
\begin{equation}
\psi_{1}=\left(1+\sum_{i=1}^{\infty}\Psi_i^{[1]}\lambda^{-i}\right)\exp{%
\left(\frac{\mathrm{i}}{4}\lambda s\right)},
\end{equation}
then we have
\begin{equation}  \label{add1}
\Psi_{1,s}^{[1]}=\frac{\mathrm{i}}{2}\sigma |q_n|^2.
\end{equation}

By Darboux transformation, it follows that
\begin{equation}
\begin{bmatrix}
\psi_1^{[1]} \\
\phi_1^{[1]} \\
\end{bmatrix}%
=\left(I+\sum_{i=1}^{\infty}T^{[i]}\lambda^{-i}\right)\left[%
\begin{bmatrix}
1 \\
0 \\
\end{bmatrix}%
+\sum_{i=1}^{\infty}\Psi_i\lambda^{-i}\right]\exp{\left(\frac{\mathrm{i}}{4}%
\lambda s\right)}.
\end{equation}
Moreover, we can obtain that
\begin{equation}  \label{add2}
\left(T^{[1]}_{1,1}\right)_s+\Psi_{1,s}^{[1]}=\frac{\mathrm{i}}{2}\sigma
|q_n^{[1]}|^2,
\end{equation}
where the element $T_{n}^{[1]}[i,j]$ denotes the $(i,j)$ element of matrix $%
T_n^{[1]}.$ Together with \eqref{add1}, we can obtain that
\begin{equation}  \label{result1}
|q_n^{[1]}|^2=|q_n|^2-2\mathrm{i}\sigma \left(T_n^{[1]}[1,1]\right)_s.
\end{equation}
Since $T_n$ is the Darboux matrix, it satisfies the following relation
\begin{equation*}
T_{n+1}U_n=U_n^{[1]}T_n.
\end{equation*}
It follows that
\begin{equation}  \label{result3}
\begin{split}
q_n^{[1]}&=q_n+T_n^{[1]}[2,1], \\
a\rho_n^{[1]}&=a\rho_n+\mathrm{i}(E-1)T_{n}^{[1]}[1,1].
\end{split}%
\end{equation}
On the other hand, since
\begin{equation*}
|y_1\rangle_s=\left(\frac{\mathrm{i}}{4}\lambda_1\sigma_3+\frac{\mathrm{i}}{2%
}Q\right)|y_1\rangle,\,\, -\langle y_1|_tJ=\langle y_1|J\left(\frac{\mathrm{i%
}}{4}\lambda_1^*\sigma_3+\frac{\mathrm{i}}{2}Q\right)
\end{equation*}
it follows that
\begin{equation}
\left(\frac{\langle y_1|J|y_1\rangle}{\lambda_1^*-\lambda_1}\right)_s=\frac{%
\mathrm{i}}{4}(-\sigma |\psi_1|^2+|\phi_1|^2)
\end{equation}
we can obtain that
\begin{equation}
\begin{split}
(E-1)T_{n}^{[1]}[1,1]&=(E-1)\left(\frac{(\lambda_1^*-\lambda_1)|\psi_{1,n}|^2%
}{\langle y_{1,n}|J|y_{1,n}\rangle}\right) \\
&=(E-1)\left(\frac{-\sigma(\lambda_1^*-\lambda_1) |\phi_{1,n}|^2}{\langle
y_{1,n}|J|y_{1,n}\rangle}\right) \\
&=(E-1)\left(\frac{(\lambda_1^*-\lambda_1)(|\psi_{1,n}|^2-\sigma|%
\phi_{1,n}|^2)}{2\langle y_{1,n}|J|y_{1,n}\rangle}\right) \\
&=2\mathrm{i}\ln_{s}\left(\frac{E(\langle y_{1,n}|J|y_{1,n}\rangle)}{\langle
y_{1,n}|J|y_{1,n}\rangle}\right).
\end{split}%
\end{equation}
Similarly, we have
\begin{equation}
(T_{n}^{[1]}[1,1])_s=2\mathrm{i}\ln_{ss}\left(\frac{\langle
y_{1,n}|J|y_{1,n}\rangle}{\lambda_1^*-\lambda_1}\right).
\end{equation}
Finally, combining the equations \eqref{result1} and \eqref{result2}, we
obtain the formulas \eqref{backlund}. This completes the proof. $\square$
%In order to drive a generalized Darboux transformation, we need the
%following lemma:
%\begin{lem}
%\end{lem}

Assume that we have $N$ different solutions $|y_{i,n}\rangle=(\psi_{i,n},%
\phi_{i,n})^T$ at $\lambda=\lambda_i$ $(i=1,2,\cdots, N)$, then we can
construct the $N$-fold DT. For simplicity, we ignore the subscript $_n$ in $%
|y_{i,n}\rangle$ and $\langle y_{i,n}|$. Furthermore, we have the following
generalized Darboux matrix

\begin{prop}
The general Darboux matrix can be represented as
\begin{equation}  \label{gDT}
T_{n,N}=I+Y M_n^{-1}D^{-1}Y^{\dag}J,
\end{equation}
where the first subscript in $T_{n,N}$ represents that the matrix is
dependent with the variable $n$, the second one represents the $N$-fold DT,
\begin{equation*}
\begin{split}
Y=&\left[|y_1^{[0]}\rangle,|y_1^{[1]}\rangle,\cdots,|y_1^{[n_1-1]}\rangle,%
\cdots, |y_r^{[0]}\rangle,|y_r^{[1]}\rangle,\cdots,|y_r^{[n_r-1]}\rangle%
\right], \\
M_n=&%
\begin{bmatrix}
M_{11} & M_{12} & \cdots & M_{1r} \\
M_{21} & M_{22} & \cdots & M_{2r} \\
\vdots & \vdots & \ddots & \vdots \\
M_{21} & M_{22} & \cdots & M_{2r}
\end{bmatrix}%
,\,\, M_{ij}=%
\begin{bmatrix}
M_{ij}^{[1,1]} & M_{ij}^{[1,2]} & \cdots & M_{ij}^{[1,n_j]} \\
M_{ij}^{[2,1]} & M_{ij}^{[2,2]} & \cdots & M_{ij}^{[2,n_j]} \\
\vdots & \vdots & \ddots & \vdots \\
M_{ij}^{[n_i,1]} & M_{ij}^{[n_i,2]} & \cdots & M_{ij}^{[n_i,n_j]}
\end{bmatrix}%
, \\
D=&\mathrm{diag}\left(D_1,D_2\cdots,D_r\right),\,\, D_i=%
\begin{bmatrix}
D_i^{[0]} & \cdots & D_i^{[n_i-1]} \\
0 & \ddots & \vdots \\
0 & 0 & D_i^{[0]}
\end{bmatrix}%
,
\end{split}%
\end{equation*}
and
\begin{equation*}
\begin{split}
|y_i(\lambda_i+\alpha_i\epsilon_i)\rangle=&\sum_{k=0}^{n_i-1}|y_i^{[k]}%
\rangle \epsilon_i^k+O(\epsilon_i^{n_i}),\,\, \frac{1}{\lambda-\lambda_i^*-%
\alpha_i\epsilon_i^*}=\sum_{k=0}^{n_i-1}D_i^{[k]}\epsilon_i^{*k}+O(%
\epsilon_i^{*n_i}) \\
\frac{\langle
y_i(\lambda_i+\alpha_i\epsilon_i)|J|y_j(\lambda_j+\alpha_j\epsilon_j)\rangle%
}{\lambda_i^*-\lambda_j+\alpha_i^*\epsilon_i^*-\alpha_j\epsilon_j}
=&\sum_{k=1}^{n_i}\sum_{l=1}^{n_j}M_{ij}^{[k,l]}\epsilon_i^{*k}%
\epsilon_j^{l}+O(\epsilon_i^{*n_i},\epsilon_j^{n_j}).
\end{split}%
\end{equation*}
The general B\"acklund transformations are
\begin{equation}  \label{gBT}
\begin{split}
\rho_n^{[N]}=&\rho_n-\frac{2}{a} \ln_{s}\left(\frac{E(\det(M_n))}{\det(M_n)}%
\right), \\
q_n^{[N]}=&q_n+\frac{\det(G_n)}{\det(M_n)}, \\
|q_n^{[N]}|^2=&|q_n|^2+4\sigma\ln_{ss}(\det(M_n))
\end{split}%
\end{equation}
where $G_n=%
\begin{bmatrix}
M & Y_1^{\dag} \\
-Y_2 & 0
\end{bmatrix}
$, $Y_k$ represents the $k$-th row of matrix $Y$.
\end{prop}

\textbf{Proof:}
Through the standard iterated step for DT, we can obtain the
$N$-fold DT
\begin{equation}  \label{n-fod-dt}
T_{n,N}=I+YM_n^{-1}D^{-1}Y^{\dag}J,
\end{equation}
where $Y=\left[|y_1\rangle,|y_2\rangle,\cdots,|y_N\rangle\right],$ and
\begin{equation*}
M_n=\left(\frac{\langle y_i|J|y_j\rangle}{\lambda_i^*-\lambda_j}%
\right)_{1\leq i,j\leq N},\,\, D=\mathrm{diag}\left(\lambda-\lambda_1^*,%
\lambda-\lambda_2^*,\cdots,\lambda-\lambda_N^*\right).
\end{equation*}
Since $T_n$ is the Darboux matrix, it satisfies the following relation
\begin{equation}
T_{n+1,N}U_n=U_n^{[N]}T_{n,N}.
\end{equation}
By using the following identities
\begin{equation}  \label{linalglem}
\begin{split}
&\phi M^{-1}\psi^{\dag}=
\begin{vmatrix}
M & \psi^{\dag} \\
-\phi & 0
\end{vmatrix}%
/|M|, \\
&1+\phi M^{-1}\psi^{\dag}=
\begin{vmatrix}
M & \psi^{\dag} \\
-\phi & 1
\end{vmatrix}%
/|M|=\frac{\det(M+\psi^{\dag}\phi)}{\det(M)}.
\end{split}%
\end{equation}
where $M$ is a $N\times N$ matrix, $\phi$, $\psi$ are a $%
1\times N$ vectors, we could derive
\begin{equation}  \label{result2}
\begin{split}
q_n^{[N]}&=q_n+T_{n,N}^{[1]}[2,1]=q_n+\frac{\det(G_n)}{\det(M_n)}, \\
a\rho_n^{[N]}&=a\rho_n+\mathrm{i}(E-1)T_{n,N}^{[1]}[1,1], \\
|q_n^{[N]}|^2&=|q_n|^2-2\mathrm{i}\sigma \left(T_{n,N}^{[1]}[1,1]\right)_s.
\end{split}%
\end{equation}
Together with the following equalities
\begin{equation}
\begin{split}
(E-1)(T_{n,N}^{[1]}[1,1])&=(E-1)\left(Y_1M^{-1}Y_1^{\dag}\right)=(E-1)%
\left(-\sigma Y_2M^{-1}Y_2^{\dag}\right) \\
&=(E-1)\left(\frac{Y_1M^{-1}Y_1^{\dag}-\sigma Y_2M^{-1}Y_2^{\dag}}{2}\right)
\\
&=2\mathrm{i}\ln_{s}\left(\frac{E\det(M_n)}{\det(M_n)}\right), \\
(T_{n,N}^{[1]}[1,1])_s&=2\mathrm{i}\ln_{ss}\left(\det(M_n)\right),
\end{split}%
\end{equation}
we can readily obtain the formula \eqref{gBT} from the above $N$-fold DT %
\eqref{n-fod-dt}. To complete the generalized DT, we set
\begin{equation*}
\begin{split}
\lambda_{r+1}&=\lambda_1+\alpha_1\varepsilon_{1,1},\,|y_{r+1}\rangle=|y_1(%
\lambda_{r+1})\rangle;\,\,\cdots,
\lambda_{r+n_1-1}=\lambda_1+\alpha_1\varepsilon_{1,n_1-1},\,|y_{r+n_1-1}%
\rangle=|y_1(\lambda_{r+n_1-1})\rangle; \\
\lambda_{r+n_1}&=\lambda_2+\alpha_2\varepsilon_{2,1},\,|y_{r+n_1}%
\rangle=|y_2(\lambda_{r+n_1})\rangle;\,\,\cdots,
\lambda_{r+n_1-1}=\lambda_2+\alpha_2\varepsilon_{2,n_2-1},\,|y_{r+n_1+n_2-2}%
\rangle=|y_2(\lambda_{r+n_1-1})\rangle; \\
&\vdots \\
\lambda_{N-n_r+1}&=\lambda_r+\alpha_r\varepsilon_{r,1},\,|y_{N-n_r+1}%
\rangle=|y_r(\lambda_{N-n_r+1})\rangle;\,\,\cdots,
\lambda_{N}=\lambda_r+\alpha_r\varepsilon_{r,n_r-1},\,|y_{N}\rangle=|y_r(%
\lambda_{N})\rangle. \\
\end{split}%
\end{equation*}
Taking limit $\varepsilon_{i,j}\rightarrow 0$, we can obtain the generalized
DT \eqref{gDT} and formulas \eqref{gBT}. $\square$

With the aid of the generalized DT, one can construct more general exact
solutions from the trivial solution of the original equation. Departing from
the zero seed solution, one-, multi-bright soliton solutions can be
constructed for the focusing CSP equation. Starting from the plane wave seed
solution, the multi-breather and high-order rogue wave solutions can be
derived for the focusing CSP equation; while one-, two- and multi-dark
soliton solutions can be obtained for the defocusing CSP equation. The
detailed results and the explicit expressions, as well as dynamics for these
solutions, are presented in the subsequent two sections.

On account of \eqref{backlund}, the coordinates transforation between $%
x_n^{[N]}$ and $x_n$ can be represented as
\begin{equation}  \label{c-tran}
x_n^{[N]}=x_n-\frac{2}{a}\ln_s(\det(M_n)),
\end{equation}
where $x_n$ represents the original coordinates. The second equation in %
\eqref{backlund} and coordinates expressions \eqref{c-tran} constitutes the
solutions for semi-discrete CSP equation \eqref{sem-csp}.

\section{Single and multi-bright solutions}
\label{section3} In this section, we construct the exact solution through
formula \eqref{gBT} as the application of DT. The general bright will be constructed for the focusing CSP equation ($\sigma=1$). To this end,
%\subsection{Single bright soliton solution and multi-soliton solution}
we start with a seed solution
\begin{equation*}
\rho_n^{[0]}=\frac{\gamma}{2},\,\, q_n^{[0]}=0,\,\, \gamma>0.
\end{equation*}
The coordinates for semi-discrete CSP \eqref{sem-csp} can be obtained
\begin{equation*}
x_n(s)=\frac{\gamma}{2} n a ,\,\, t=-s.
\end{equation*}
Solving the Lax pair equation \eqref{cd-lax} with $(\rho_n,q_n;\lambda)=(%
\rho_n^{[0]},q_n^{[0]};\lambda_i=\alpha_i+\mathrm{i}\beta_i)$, $\beta_i>0$,
one obtains a special solution
\begin{equation}  \label{thetai}
|y_{i,n}\rangle=%
\begin{bmatrix}
\mathrm{e}^{\theta_{i,n}} \\
1 \\
\end{bmatrix}%
,\,\,\theta_{i,n}=\frac{\mathrm{i}}{2}\lambda_i s+n\ln\left(\frac{\lambda_i-%
\frac{\mathrm{i}a\gamma}{2}}{\lambda_i+\frac{\mathrm{i}a\gamma}{2}}%
\right)+a_i,
\end{equation}
where $a_i$s are complex parameters. Then we can obtain that the single
soliton solution through the formula \eqref{gBT}:
\begin{equation*}
\begin{split}
\rho_n^{[1]}=&\frac{\gamma}{2}+\frac{\beta_1}{a}[\tanh(\theta_{1,n+1}^{R})-
\tanh(\theta_{1,n}^{R})]>0, \\
q_n^{[1]}=&\beta_1\mathrm{sech}(\theta_{1,n}^{R})\mathrm{e}^{-\mathrm{i}%
\theta_{1,n}^{I}-\frac{\pi\mathrm{i}}{2}}, \\
x_n^{[1]}=&\frac{\gamma}{2} na+\frac{\beta_1}{a}\tanh(\theta_{1,n}^{R}),\,%
\, t=-s,
\end{split}%
\end{equation*}
where the superscripts $^{R}$, $^{I}$ represent the real part and imaginary part, respectively,
\begin{equation*}
\theta_{1,n}^{R}=-\frac{\beta_{{1}}}{2}s+\frac{n}{2}g_1+a_1^{R},\,\,%
\theta_{1,n}^{I}=\frac{1}{2}\alpha_1s+n\mathrm{arg}\left(\frac{\lambda_1-%
\frac{\mathrm{i}a\gamma}{2}}{\lambda_1+\frac{\mathrm{i}a\gamma}{2}}%
\right)+a_1^{I},\,\, g_1=\ln \left( {\frac{4{\alpha_{1}}^{2}+ \left(2\beta_{{%
1}}-a{\gamma}\right)^{2}}{4{\alpha_{{1}}}^{ 2}+ \left( 2\beta_{{1}}+a{\gamma}
\right) ^{2}}} \right),
\end{equation*}
where $4{\alpha_{1}}^{2}+ \left(2\beta_{{1}}-a{\gamma}\right)^{2}\neq 0.$
The soliton $|q_n^{[1]}|^2$ propagates along the line $\theta_{1,n}^{R}=0.$
The peak values $|q_n^{[1]}|_{max}^2=\beta_1^2$ locate at $(x,t)=(n,\frac{1}{%
\beta_1}(ng_1+2a_1^{R})).$ To obtain the smooth bright soliton for semi-discrete
CSP equation, we require that $\rho_n^{[1]}>0$ for all $n\in \mathbf{Z}$ and
$t\in \mathbf{R}$.
%i.e. $\frac{1}{2}+\frac{2\beta_1}{a\gamma}\tanh(\frac{1}{4%
%}g_1)>0$ or $0<\frac{2\beta_1}{a\gamma}<\sqrt{\frac{1}{2}+4\alpha_1^2/(a%
%\gamma)^2}$.
Otherwise, the bright soliton will be either cusp or soliton solution.
%can be derived, but
%they are not the functions in general.

{{Inserting the equation \eqref{thetai} into formula \eqref{gBT},}}, we can deduce the multi-bright soliton solution as follows:
\begin{equation}  \label{nsoliton}
\begin{split}
\rho_n^{[N]}&=\frac{\gamma}{2}-\frac{2}{a} \ln_{s}\left(\frac{\det(M_{n+1})}{%
\det(M_n)}\right) \\
q_n^{[N]}&=\frac{\det(G_n)}{\det(M_n)}, \\
x_n^{[N]}&=\frac{\gamma}{2}  n a-\frac{2}{a}\ln_{s}\det(M_n),\,\, t=-s,
\end{split}%
\end{equation}
where
\begin{equation*}
\begin{split}
M_n&=\left(\frac{\mathrm{e}^{\theta_{i,n}^*+\theta_{j,n}}+1}{%
\lambda_i^*-\lambda_j}\right)_{1\leq i,j\leq N},\,\, G_n=%
\begin{bmatrix}
M_n & Y_{1,n}^{\dag} \\
-Y_{2,n} & 0
\end{bmatrix}
, \\
Y_{1,n}&=%
\begin{bmatrix}
\mathrm{e}^{\theta_{1,n}}, & \mathrm{e}^{\theta_{2,n}}, & \cdots, & \mathrm{e%
}^{\theta_{N,n}}
\end{bmatrix}%
, \,\, Y_{2,n}=%
\begin{bmatrix}
1, & 1, & \cdots, & 1
\end{bmatrix},
\end{split}%
\end{equation*}
the expression for $\theta_{i,n}$ is given in \eqref{thetai}. Here we require $\rho_n^{[N]}>0$ such that the solutions are well-posed.

In what follows, we will prove that the multi-bright soliton solution of semi-discrete CSP equation converges to the multi-bright soliton solution obtained in \cite{LFZPhysD,FengShen_ComplexSPE}. Referring to the Taylor expansion
\begin{equation}\label{Taylor}
\ln(1\pm x)= \pm x + o(x^2)\,,
\end{equation}
we have
\begin{equation}\label{brghtdiscon}
    \begin{split}
     n\ln\left(\frac{\lambda_i-\frac{\mathrm{i}a\gamma}{2}}{\lambda_i+\frac{\mathrm{i}a\gamma}{2}}\right)
     & \approx -n \frac{\mathrm{i}a\gamma}{\lambda_i}\\
     &=-\frac{\mathrm{i}\gamma}{\lambda_i}y
    \end{split}
\end{equation}
by letting $na=y$ in the continuous limit $a\ to 0$. Therefore $\theta_{i,n}$ agrees with (32) in \cite{LFZPhysD} by noticing the correspondence $\theta_{i,n} \to 2\theta_{i,n}$ and $\gamma \to -\gamma$.

%{\bf {Here I will prove in the continuous limit $a \to 0$, the multi-bright soliton solution will approach to the continuous case we obtained previously}}

\begin{figure}[tbh]
\centering
\subfigure[Single bright
soliton]{\includegraphics[height=50mm,width=65mm]{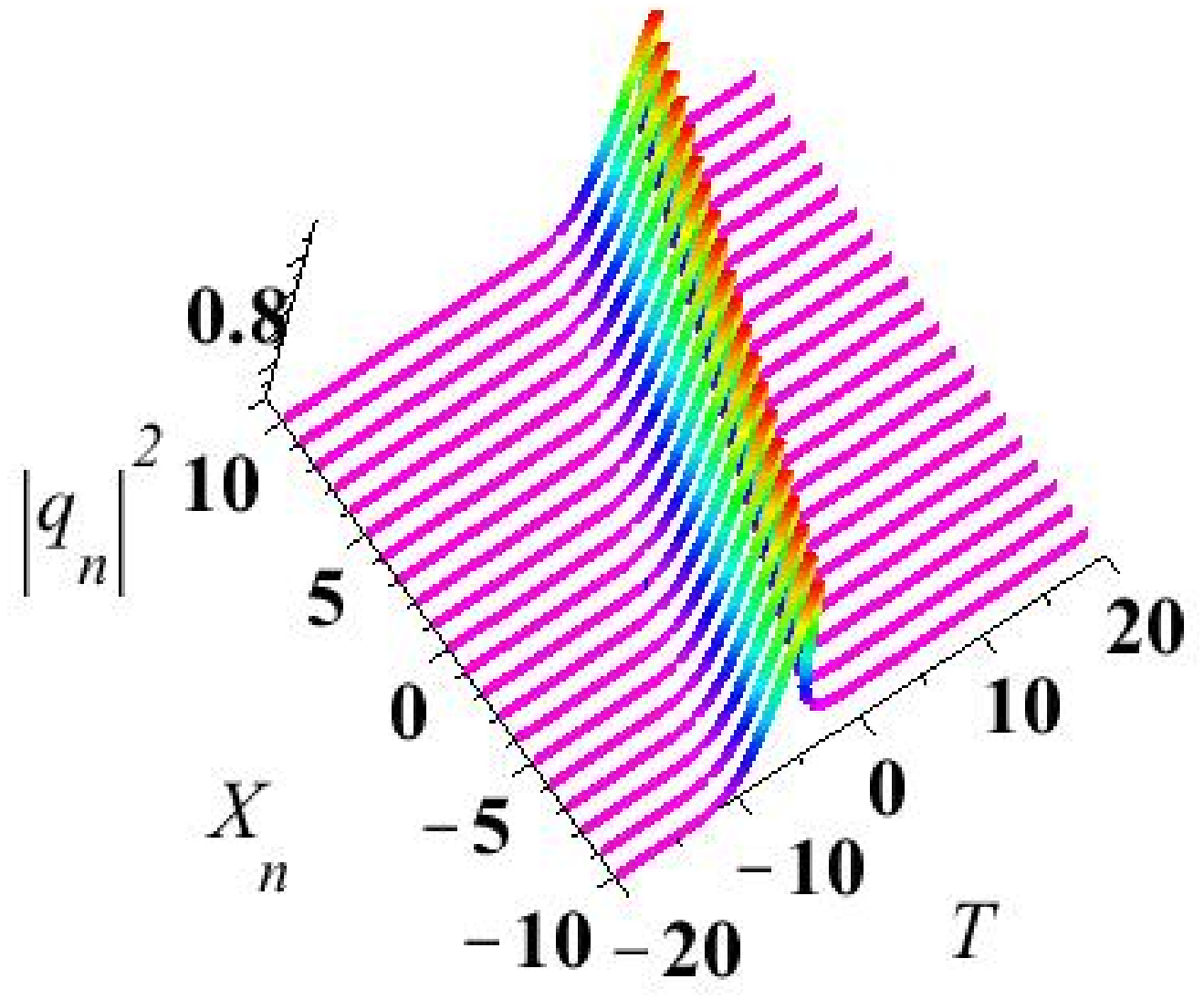}} \hfil
\subfigure[Two bright
soliton]{\includegraphics[height=50mm,width=65mm]{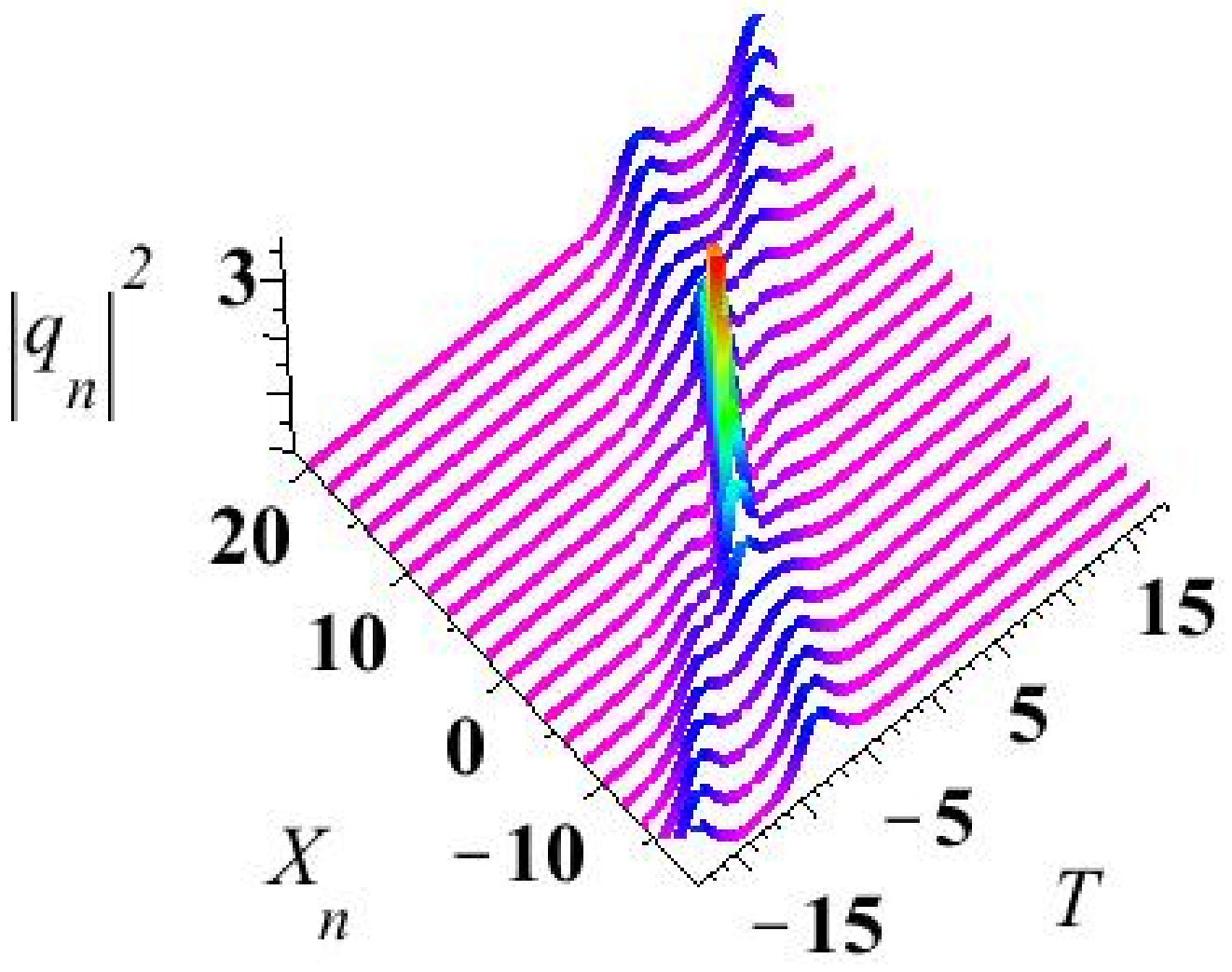}}
\caption{(color online): Bright solitons}
\label{figure1}
\end{figure}

In particular, we give two soliton solution explicitly through the above general formula \eqref{nsoliton}:
\begin{equation}\label{2soliton}
    \begin{split}
     \rho_n^{[2]}&=\frac{\gamma}{2}-\frac{2}{a} \ln_{s}\left(\frac{\det(M_{n+1}^{[2]})}{\det(M_n^{[2]})}\right) \\
      q_n^{[2]}&=\frac{\det(G_n^{[2]})}{\det(M_n^{[2]})}, \\
      x_n^{[2]}&=\frac{\gamma}{2} n a -\frac{2}{a}\ln_{s}\det(M_n^{[2]}),\,\, t=-s,
    \end{split}
\end{equation}
where
\begin{equation*}
\begin{split}
M_n^{[2]}&={\frac { \left( {{\rm e}^{\theta_{1,n}+\theta_{1,n}^*}
}+1 \right)  \left( {{\rm e}^{\theta_{2,n}+\theta_{2,n}^*}}
+1 \right) }{4\beta_1\,\beta_2}}-{\frac { \left( {{\rm e}^{\theta_{2,n}+\theta_{1,n}^*}}+1
\right)  \left( {{\rm e}^{
{\theta_{1,n}}+\theta_{2,n}^*}}+1 \right) }{ \left( -\alpha_1+\alpha_2 \right)^{2}+
\left( \beta_1+\beta_2
 \right) ^{2}}},  \\
   G_n^{[2]}&=\left( {\frac {{{\rm e}^{\theta_{2,n}+\theta_{1,n}^*}}+1 }{(\beta_1+\beta_2)+(\alpha_1-\alpha_2){\rm i}}}-{\frac {{{\rm e}^{\theta_{1,n}+
\theta_{1,n}^*
}}+1}{2\beta_1}} \right) {{\rm e}^{\theta_{2,n}^*}}+
 \left({\frac {{{\rm e}^{\theta_{1,n}+\theta_{2,n}}}+1}{(\beta_1+\beta_2)+{\rm i}(\alpha_2-\alpha_1)}}-{\frac {{{\rm e}^{\theta_{2,n}+\theta_{2,n}^*}
}+1}{2\beta_2}}\right)
{{\rm e}^{\theta_{1,n}^*}}.
\end{split}
\end{equation*}
%Using the computer soft, we could exhibit the dynamics for the two or more soliton solution. For instance, if
The single bright soliton is illustrated
in Fig. \ref{figure1}(a) with the parameters  $\gamma=1$, $a=2$,
$\alpha_1=2$, $\beta_1=1$, $a_1=0,$. To exhibit the dynamics for the two-soliton solution as shown in Fig. \ref{figure1} (b)), we choose the parameters $\gamma=2$, $a=2$, $\alpha_1=2$, $\beta_1=1$, $a_1=0,$ $\alpha_2=1$, $\beta_2=1$, $a_2=0$. It
is seen that the two solitons interact with each other elastically.
\section{Single and multi-dark soliton solution}
\label{section4}

In this section, we construct the one- and multi-dark
soliton solution for the defocusing CSP equation ($\sigma=-1$) in detail.
Generally, the DT cannot apply to derive the dark solitons directly, since
the spectral points of dark solitons locate in the real axis and the Darboux
matrix is trivial if $\lambda_1=\lambda_1^*$. The authors in \cite{Ling}
develop a method to yield the dark soliton and multi-dark solitons through
the Darboux transformation together with the limit technique. In what
followings, we follow the steps in reference \cite{Ling} to give the dark
and multi-dark solitons for the semi-discrete CSP equation.

The dark solution and multi-dark solution can be constructed from the seed solution--plane wave solution through formula \eqref{gBT}.
We depart from the seed solution
\begin{equation}\label{seed2}
    \rho_n^{[0]}=\frac{\gamma}{2},\,\, q_n^{[0]}=\frac{\beta}{2}{\rm e}^{{\rm i}\theta_n},\,\, \theta_n=bn+\frac{c}{2}s,\,\, c=\frac{a\gamma}{2}\frac{\sin(b)}{\cos(b)-1},\,\, \gamma>0,\,\,\beta\geq0,\,\,\, b\neq k\pi,\,\, k\in\mathbb{Z}.
\end{equation}
Then we have the solution vector for Lax pair equation \eqref{cd-lax} with $(q_n,\rho_n;\lambda)=(q_n^{[0]},\rho_n^{[0]};\lambda_i)$,
\begin{equation}\label{sol-vec}
    |y_{i,n}\rangle=KL_iE_i=K\begin{bmatrix}
                               \widehat{\phi_{i,n}} \\
                               \beta\widehat{\psi_{i,n}} \\
                             \end{bmatrix}
    ,\,\, K=\mathrm{diag}\left({\rm e}^{-\frac{{\rm i}}{2}\theta_n},{\rm e}^{\frac{{\rm i}}{2}\theta_n}\right),\,\,\lambda_i\neq -c+{\rm i}\beta,
\end{equation}
where $|y_{i,n}\rangle$ discards a function,
\begin{equation*}
    L_i=\begin{bmatrix}
          1 & 1 \\[10pt]
          {\displaystyle \frac{\beta}{c+\chi_{i}^{+}}} & {\displaystyle \frac{\beta}{c+\chi_{i}^{-}}}  \\
        \end{bmatrix},\,\, E_i=\begin{bmatrix}
                                  {\rm e}^{\omega_{i,n}}\\
                                  \alpha_i(\bar{\lambda_i}-\lambda_i){\rm e}^{-\omega_{i,n}}\\
                               \end{bmatrix}
\end{equation*}
and
\begin{equation}\label{theata}
\begin{split}
\omega_{i,n}=&\frac{{\rm i}}{4}\xi_i s+\frac{n}{2}\ln  \left( {\frac {\sin(\frac{b}{2})\left(\frac{1}{2}{\rm i}a \gamma-
\xi_i \right) +{\rm i}\cos(\frac{b}{2}) \lambda_i}{
\sin(\frac{b}{2})  \left( \frac{1}{2}{\rm i}a{\gamma}+\xi_i\right) +{\rm i}
\cos(\frac{b}{2}) \lambda_i}} \right)
+a_i,\\
\chi_{i}^{\pm}=&\lambda_i\pm\xi_i,\,\, \xi_i=\sqrt{(\lambda_i+c)^2-\beta^2}.
\end{split}
\end{equation}
$\alpha_i$s are appropriate complex parameters and $a_i$s are real parameters.
In order to derive the single dark soliton solution, we consider only $\lambda_1$ and replace the parameter condition \eqref{theata} with
\begin{equation}
\chi_{1}^{\pm}=\beta[\cos(\varphi_1)\pm {\rm i}\sin(\varphi_1)]\,\,, \xi_1={\rm i} \beta\sin(\varphi_1).
\end{equation}
and $0<\varphi_i<\pi,$ $a_i\in \mathbf{R}$.
By taking a limit process $\lambda _{1}\rightarrow \bar{%
\lambda }_{1}$ similar to the one in \cite{Ling}, the single dark soliton solution can be obtained as follows
\begin{equation}
    \begin{split}
       \rho_n^{[1]}=&\frac{\gamma}{2}+\frac{\beta}{2a}\sin(\varphi_1)(\tanh(Z_{1,n+1})-\tanh(Z_{1,n})),  \\
       q_n^{[1]}=&\frac{\beta}{2}\left[1-{\rm i}\sin(\varphi_1){\rm e}^{-{\rm i}\varphi_1}-{\rm i}\sin(\varphi_1){\rm e}^{-{\rm i}\varphi_1}\tanh(Z_{1,n})\right]{\rm e}^{{\rm i}\theta_n},\\
        x_n^{[1]}=&\frac{\gamma}{2} a n+\frac{\beta^2}{8}s+\frac{\beta}{2a}\sin(\varphi_1)\tanh(Z_{1,n}),\,\, t=-s,
    \end{split}
\end{equation}
where
\begin{equation}
\label{Z1n}
Z_{1,n}(\varphi _{1},a_{1})=\frac{n}{2}\ln \left( \frac{a\gamma +2\beta \sin
(\frac{1}{2}b)\cos (\frac{1}{2}b+\varphi _{1})}{a\gamma +2\beta \sin (\frac{1%
}{2}b)\cos (\frac{1}{2}b-\varphi _{1})}\right) -\frac{\beta }{4}\sin
(\varphi _{1})s+a_{1}.
\end{equation}%
and $a_{1}$ is a real parameter.

Next, we proceed to finding $N$-dark soliton solution.
Based on the $N$-soliton solution \eqref{gBT} to the defocusing semi-discrete CSP equation, it then follows
\begin{equation}
\begin{split}
q_n[N]=& \frac{\beta }{2}\left[ 1+\widehat{Y_{2,n}}M_n^{-1}\widehat{Y_{1,n}}^{\dag }%
\right] \mathrm{e}^{\mathrm{i}\theta_n }=\frac{\beta }{2}\left[ \frac{\det (H_n)%
}{\det (M_n)}\right] \mathrm{e}^{\mathrm{i}\theta_n }, \\
x_n=& \frac{\gamma }{2}an+\frac{\beta^{2}}{8}s-\frac{2}{a}\ln _{s}(\det (M_n)),\quad \,t=-s,
\end{split}
\label{nsoliton1}
\end{equation}
where
\begin{equation*}
\begin{split}
M_n& =\left( \frac{\langle y_{i,n}|\sigma _{3}|y_{j,n}\rangle }{2(\bar{\lambda}_{i}-\lambda _{j})}\right)_{1\leq i,j\leq N},\quad \,H_n=M_n+Y_{1,n}^{\dag
}Y_{2,n}, \\
\widehat{Y_{1,n}}& =\left[ \widehat{\phi _{1,n}},\widehat{\phi _{2,n}},\cdots ,%
\widehat{\phi _{N,n}}\right] ,\quad \,\widehat{Y_{2,n}}=\left[ \widehat{\psi _{1,n}},%
\widehat{\psi _{2,n}},\cdots ,\widehat{\psi _{N,n}}\right] .
\end{split}%
\end{equation*}
In general, the above $N$-soliton solution \eqref{nsoliton} is singular.
In order to derive the $N$-dark soliton solution through the DT
method, we need to take a limit process $\lambda _{i}\rightarrow \bar{%
\lambda }_{i},(i=1,2,\cdots ,N)$.
By a tedious procedure which is omitted here, we finally have the
$N$-dark soliton solution to the defocusing semi-discrete CSP equation
\eqref{sem-csp} as follows
%In summary, we finally have the $N$-dark soliton solution to the defocusing
%semi-discrete CSP equation (\ref{sem-csp}) as follows
\begin{prop}
\begin{equation} \label{ndark1}
\begin{split}
\rho _{n}^{[N]}=& \frac{\gamma }{2}-\frac{2}{a}\ln _{s}\frac{\det (G_{n+1})}{%
\det (G_{n})}, \\
q_{n}^{[N]}=& \frac{\beta }{2}\left[ \frac{\det (H_{n})}{\det (G_{n})}\right]
\mathrm{e}^{\mathrm{i}\theta _{n}}, \\
x_{n}^{[N]}=& \frac{\gamma }{2}an+\frac{\beta ^{2}}{8}s-\frac{2}{a}\ln
_{s}\det (G_{n}),\,\,t=-s,
\end{split}%
\end{equation}%
where $G_{n}=(g_{i,j})_{1\leq i,j\leq N}$, $H_{n}=(h_{i,j})_{1\leq i,j\leq
N} $,
\begin{equation} \label{n-dark-entry}
\begin{split}
g_{i,j}& =\frac{\delta _{ij}+\mathrm{e}^{Z_{i,n}+Z_{j,n}}}{\exp (-\mathrm{i}%
\varphi _{i})-\exp (\mathrm{i}\varphi _{j})}, \\
h_{i,j}& =\frac{\delta _{ij}+\mathrm{e}^{(Z_{i,n}-\mathrm{i}\varphi
_{i})+(Z_{j,n}-\mathrm{i}\varphi _{j})}}{\exp (-\mathrm{i}\varphi _{i})-\exp
(\mathrm{i}\varphi _{j})},
\end{split}%
\end{equation}%
$Z_{i,n}=Z_{1,n}(\varphi _{i},a_{i})$ and $\delta _{i,j}$ is the standard
Kronecker delta.
\end{prop}

%{\bf {Here I will prove in the continuous limit $a \to 0$, the multi-dark soliton solution will approach to the continuous case we obtained previously}}

\begin{figure}[tbh]
\centering
\subfigure[Single dark
soliton]{\includegraphics[height=50mm,width=65mm]{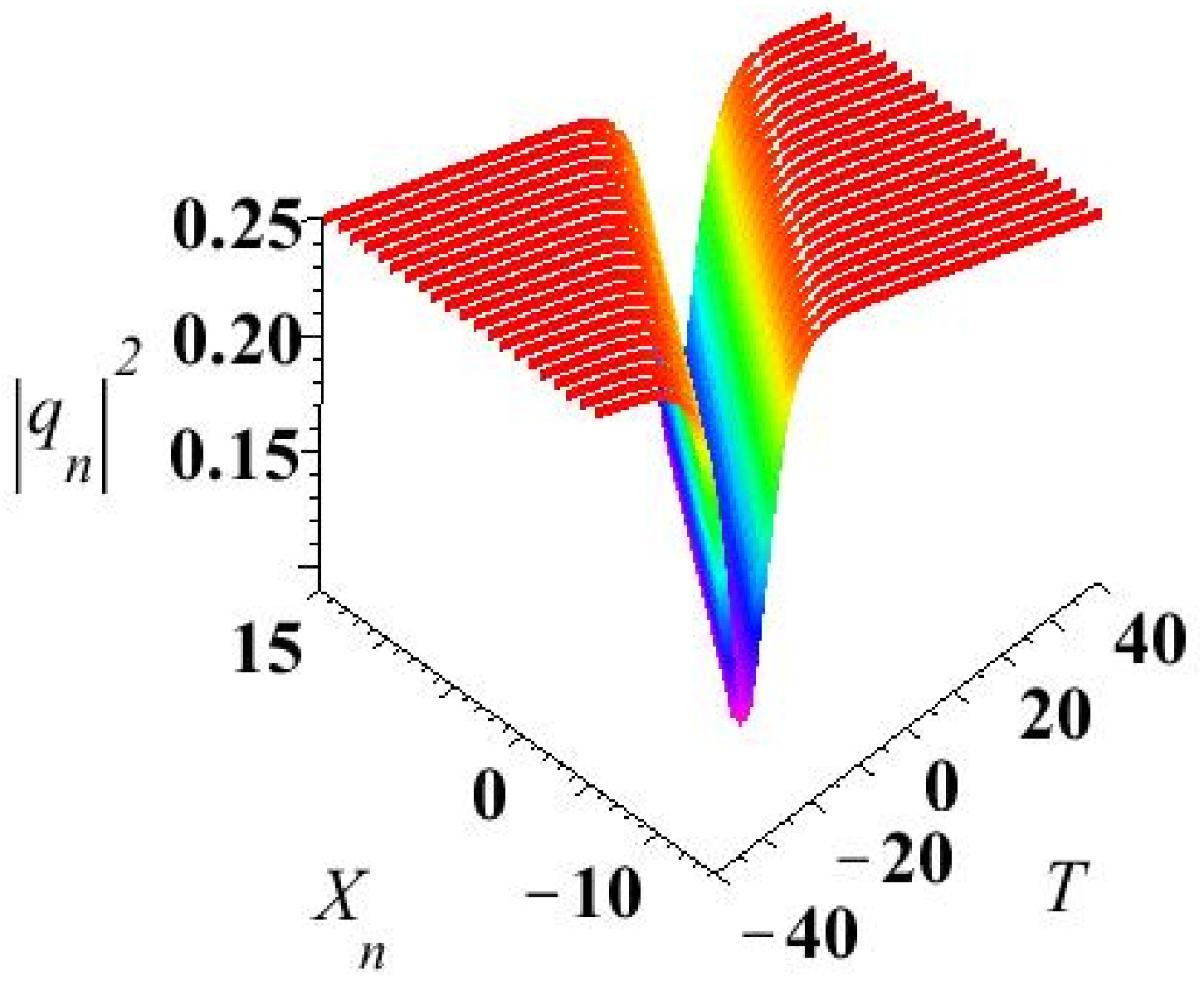}} \hfil
\subfigure[Two dark
soliton]{\includegraphics[height=50mm,width=65mm]{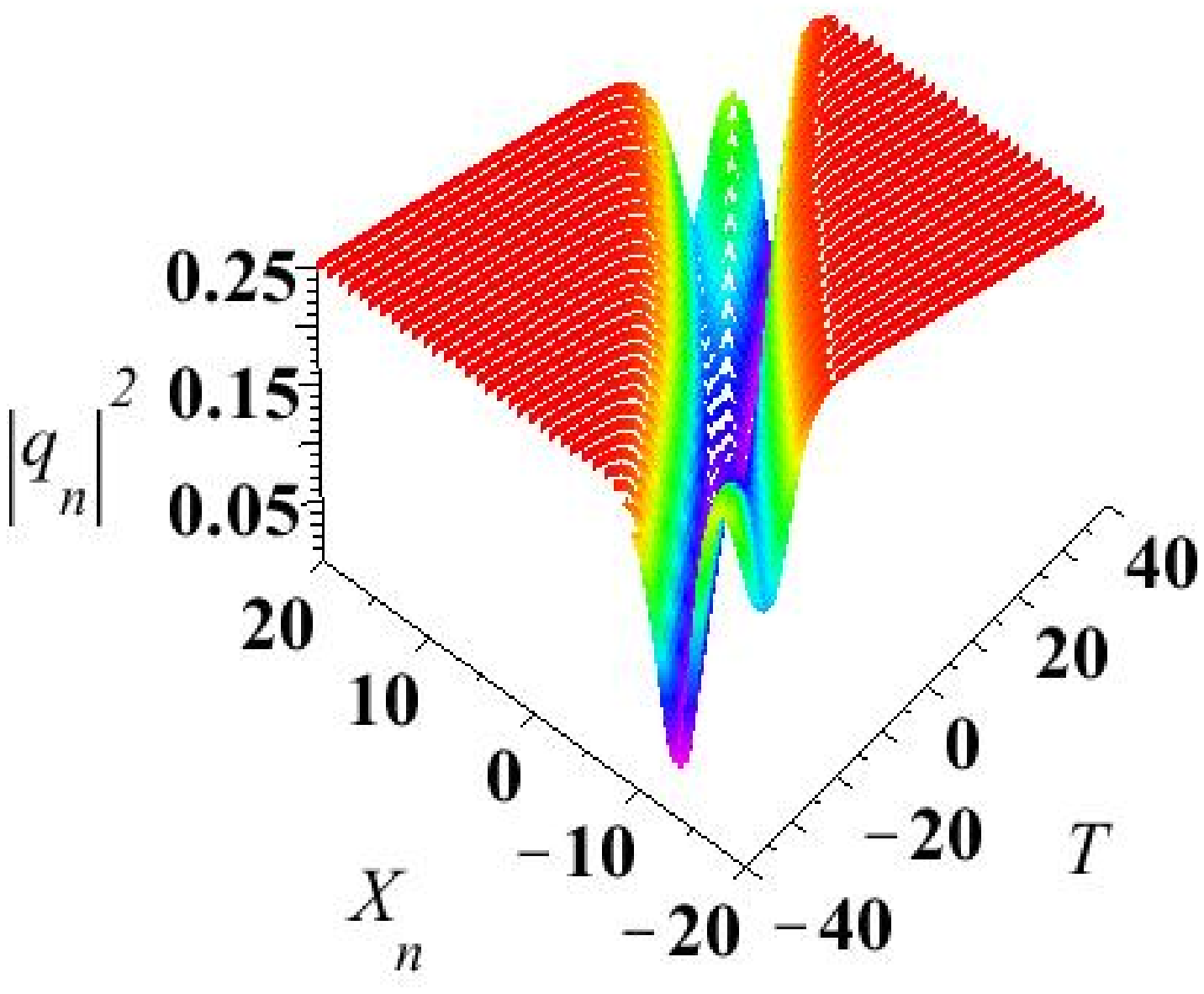}}
\caption{(color online): Dark soliton.}
\label{figure4}
\end{figure}

%\bigskip LIMNG: PLEASE ADD A DETAILED PROCESS IN OBTAINING MULTI-DARK
%SOLITON SOLUTION
By taking $N=2$ in (\ref{ndark1}) and (\ref{n-dark-entry}), the determinants
corresponding to two-dark soliton solution can be calculated as
\begin{equation}
|G_{n}|=1+e^{2Z_{i,n}}+e^{2Z_{2,n}}+a_{12}e^{2(Z_{1,n}+Z_{2,n})}\,,
\end{equation}%
\begin{equation}
|H_n|=1+e^{2(Z_{i,n}-\mathrm{i}\varphi _{1})}+e^{2(Z_{2,n}-\mathrm{i}\varphi
_{2})}+a_{12}e^{2(Z_{1,n}+Z_{2,n}-\mathrm{i}\varphi _{1}-\mathrm{i}\varphi
_{2})}\,,
\end{equation}%
where
\begin{equation}
a_{12}=\frac{\sin ^{2}\left( \frac{\varphi _{2}-\varphi _{1}}{2}\right) }{%
\sin ^{2}\left( \frac{\varphi _{2}+\varphi _{1}}{2}\right) }\,.
\end{equation}
Asymptotic analysis can be easily performed for two-soliton
interaction, which shows that the collision is always elastic. We shows an
example of such two-soliton collision. If we choose the parameter $a=1$, $%
\varphi _{1}=\arcsin (4/5)$ and $a_{1}=0$, then the single dark soliton are
shown in Figure \ref{figure4} (a). To shown their interaction for two dark
solitons, choosing the parameters $a=1$, $\varphi _{1}=\arcsin (4/5)$, $%
\varphi _{2}=\pi /2$, $a_{1}=a_{2}=0$, we arrive at the dynamics of two dark solitons (Fig. \ref{figure4}(b)), which show that the interaction between them is elastic.

Prior to the closing of this section, let us prove that the multi-dark solution converges to the its counterpart of the continuous CSP equation obtained in \cite{FenglingzhuPRE}. In the continuous limit, we assume $a=b \to 0$, it then follows $c \to -\gamma$.
Referring to the Taylor expansion \eqref{Taylor}, $Z_{i,n}$ turns out to be
\begin{equation} \label{darkdiscon}
    \begin{split}
&  \frac{n}{2}\ln \left( \frac{a\gamma +2\beta \sin
(\frac{1}{2}b)\cos (\frac{1}{2}b+\varphi _{i})}{a\gamma +2\beta \sin (\frac{1%
}{2}b)\cos (\frac{1}{2}b-\varphi _{i})}\right) -\frac{\beta }{4}\sin
(\varphi _{i})s+a_{i}
\\
& \approx n \frac{-2\beta \sin^2 (\frac{b}{2}) \sin
(\varphi _{i})} {2\beta \cos (\frac{b}{2}) \sin (\frac{b}{2})\cos
(\varphi _{i}) + a\gamma} -\frac{\beta }{4}\sin
(\varphi _{i})s+a_{i}\,.
    \end{split}
\end{equation}
Note that, between the present paper and \cite{FenglingzhuPRE}
$\gamma \to -\gamma$.
%thus, $a\gamma$ should be $-2\gamma$.
As a result $  Z_{i,n} \to \omega_i$ in \cite{FenglingzhuPRE} by letting $nb=y$ and the proof is complete.

\section{Single breather and multi-breather solutions}
\label{section5}
%\subsection{Single breather solution}
The single breather and multi-breather solution for the focusing semi-discrete CSP equation \eqref{sem-csp} ($\sigma=1$) can be constructed from the seed solution--plane wave solution through formula \eqref{gBT}. We
depart from the seed solution
\begin{equation}
\rho _{n}^{[0]}=\frac{\gamma }{2},\,\,q_{n}^{[0]}=\frac{\beta }{2}\mathrm{e}%
^{\mathrm{i}\theta _{n}},\,\,\theta _{n}=bn+\frac{c}{2}s,\,\,c=\frac{a\gamma
}{2}\frac{\sin (b)}{\cos (b)-1},\,\,\gamma >0,\,\,\beta \geq 0.
\label{seed21}
\end{equation}%
The coordinates for semi-discrete CSP \eqref{sem-csp} can be obtained
\begin{equation*}
x_{n}(s)=\frac{\gamma }{2} na-\frac{\beta ^{2}}{8}s,\,\,t=-s.
\end{equation*}%
Then we have the solution vector for Lax pair equation \eqref{cd-lax} ($\sigma=1$) with%
$(q_{n},\rho _{n};\lambda )=(q_{n}^{[0]},\rho _{n}^{[0]};\lambda _{1})$,
\begin{equation}
|y_{1,n}\rangle =KL_{1}E_{1},\,\,K=\mathrm{diag}\left( \mathrm{e}^{-\frac{%
\mathrm{i}}{2}\theta _{n}},\mathrm{e}^{\frac{\mathrm{i}}{2}\theta
_{n}}\right) ,\,\,\lambda _{1}\neq -c+\mathrm{i}\beta ,  \label{sol-vec-1}
\end{equation}%
where
\begin{equation*}
L_{1}=%
\begin{bmatrix}
1 & 1 \\[10pt]
{\displaystyle\frac{\beta }{c+\eta _{1}}} & {\displaystyle\frac{\beta }{%
c+\chi _{1}}}
\end{bmatrix}%
,\,\,E_{i}=%
\begin{bmatrix}
\mathrm{e}^{\theta _{1,n}} \\
1 \\
\end{bmatrix}%
\end{equation*}%
and
\begin{equation}
\begin{split}
\theta _{1,n}=& \frac{\mathrm{i}}{2}\xi _{1}s+n\ln \left( {\frac{\sin (\frac{%
b}{2})\left( \frac{1}{2}\mathrm{i}a\gamma -\xi _{1}\right) +\mathrm{i}\cos (%
\frac{b}{2})\lambda _{1}}{\sin (\frac{b}{2})\left( \frac{1}{2}\mathrm{i}a{%
\gamma }+\xi _{1}\right) +\mathrm{i}\cos (\frac{b}{2})\lambda _{1}}}\right)
+a_{1}, \\
\eta _{1}=& \lambda _{1}+\xi _{1},\,\,\chi _{1}=\lambda _{1}-\xi
_{1},\,\,\xi _{1}=\sqrt{\beta ^{2}+(\lambda _{1}+c)^{2}}.
\end{split}
\end{equation}
The single breather solution can be constructed from the formula %
\eqref{gBT} with the technique as in reference \cite{Ling}:
\begin{equation}
\begin{split}
\rho _{n}^{[1]}=& \frac{\gamma }{2}-\frac{2}{a}\ln _{s}\left( \frac{\cosh
(\theta _{1,n+1}^{R})\cosh (\varphi _{1}^{R}/2)-\sin (\theta
_{1,n+1}^{I})\sin (\varphi _{1}^{I}/2)}{\cosh (\theta _{1,n}^{R})\cosh
(\varphi _{1}^{R}/2)-\sin (\theta _{1,n}^{I})\sin (\varphi _{1}^{I}/2)}%
\right) >0, \\
q_{n}^{[1]}=& \frac{\beta }{2}\left[ \frac{\cosh (\theta _{1,n}^{R}-\mathrm{i%
}\varphi _{1}^{I})\cosh (\varphi _{1}^{R}/2)+\sin (\theta _{1,n}^{I}+\mathrm{%
i}\varphi _{1}^{R})\sin (\varphi _{1}^{I}/2)}{\cosh (\theta _{1,n}^{R})\cosh
(\varphi _{1}^{R}/2)-\sin (\theta _{1,n}^{I})\sin (\varphi _{1}^{I}/2)}%
\right] \mathrm{e}^{\mathrm{i}\theta _{n}}, \\
x_{n}^{[1]}=& \frac{\gamma }{2}na-\frac{\beta ^{2}}{8}s-\frac{2}{a}\ln
_{s}\left( \cosh (\theta _{1,n}^{R})\cosh (\varphi _{1}^{R}/2)-\sin (\theta
_{1,n}^{I})\sin (\varphi _{1}^{I}/2)\right) ,\,\,t=-s,
\end{split}%
\end{equation}%
where
\begin{equation*}
\xi _{i}=\beta \cosh \left[ \frac{1}{2}(\varphi _{i}^{R}+\mathrm{i}\varphi
_{i}^{I})\right] ,\,\,\eta _{i}+c=\beta \mathrm{e}^{\frac{1}{2}(\varphi
_{i}^{R}+\mathrm{i}\varphi _{i}^{I})},\,\,\chi _{i}+c=-\beta \mathrm{e}^{-%
\frac{1}{2}(\varphi _{i}^{R}+\mathrm{i}\varphi _{i}^{I})}.
\end{equation*}%
\begin{equation*}
\begin{split}
\theta _{1,n}^{R}& =\frac{\ln (g_{1})}{2}n-\frac{\beta }{2}\sinh \left(
\frac{\varphi _{1}^{R}}{2}\right) \sin \left( \frac{\varphi _{1}^{I}}{2}%
\right) s-\varphi _{1}^{R}+a_{1}^{R}, \\
\theta _{1,n}^{I}& =h_{1}n+\frac{\beta }{2}\cosh \left( \frac{\varphi
_{1}^{R}}{2}\right) \cos \left( \frac{\varphi _{1}^{I}}{2}\right) s-\varphi
_{1}^{I}+a_{1}^{I},
\end{split}%
\end{equation*}%
and
\begin{equation*}
\begin{split}
g_{1}& =\frac{\beta ^{2}\cosh ^{2}\left( \varphi _{1}^{R}/2\right) \sin
^{2}\left( b/2+\varphi _{1}^{I}/2\right) +\left[ \beta \sinh \left( \varphi
_{1}^{R}/2\right) \cos \left( b/2+\varphi _{1}^{I}/2\right) +\frac{a\gamma }{%
2\sin \left( b/2\right) }\right] ^{2}}{\beta ^{2}\cosh ^{2}\left( \varphi
_{1}^{R}/2\right) \sin ^{2}\left( b/2-\varphi _{1}^{I}/2\right) +\left[
\beta \sinh \left( \varphi _{1}^{R}/2\right) \cos \left( b/2-\varphi
_{1}^{I}/2\right) +\frac{a\gamma }{2\sin \left( b/2\right) }\right] ^{2}}, \\
h_{1}& =\mathrm{arg}\left( {\frac{\sin (\frac{b}{2})\left( \frac{1}{2}%
\mathrm{i}a\gamma -\beta \cosh \left[ \frac{1}{2}(\varphi _{1}^{R}+\mathrm{i}%
\varphi _{1}^{I})\right] \right) +\mathrm{i}\cos (\frac{b}{2})(\beta \sinh %
\left[ \frac{1}{2}(\varphi _{1}^{R}+\mathrm{i}\varphi _{1}^{I})\right] -c)}{%
\sin (\frac{b}{2})\left( \frac{1}{2}\mathrm{i}a{\gamma }+\beta \cosh \left[
\frac{1}{2}(\varphi _{1}^{R}+\mathrm{i}\varphi _{1}^{I})\right] \right) +%
\mathrm{i}\cos (\frac{b}{2})(\beta \sinh \left[ \frac{1}{2}(\varphi _{1}^{R}+%
\mathrm{i}\varphi _{1}^{I})\right] -c)}}\right) .
\end{split}%
\end{equation*}
Specially, if we choose the parameters such that $\beta =\gamma =1$, $a=2$, $%
b=\frac{\pi }{2}$, $\varphi _{1R}=0$, $\varphi _{1I}=\arcsin (\frac{3}{5})$,
$a_{1}=0$, we can obtain the explicit dynamics (Fig. \ref{figure2}(a)) for
the breather solution which is periodical in time and localized in space and
usually is called the Kuznetsov-Ma (K-M) breather.
%
%\subsection{Multi-breather and higher order rogue wave solution}

Furthermore, by using the generalized $N$-fold DT, we drive the $N$-breather solution through the formula \eqref{gBT} and some tedious
algebraic calculations as the following proposition
%\begin{equation*}
%\frac{2(\lambda _{m}^{\ast }-\lambda _{k})}{\eta _{m}^{\ast }-\eta %_{k}}=1+%
%\frac{\beta ^{2}}{(\eta _{m}^{\ast }+c)(\eta _{k}+c)},
%\end{equation*}%
%the multi-AB solution can be constructed as the following proposition:
\begin{prop}
\label{propv4} The multi-breather solution for semi-discrete CSP equation %
\eqref{sem-csp} can be represented as
\begin{equation}
\begin{split}
\rho _{n}^{[N]}& =\frac{\gamma }{2}-\frac{2}{a}\ln _{s}\left( \frac{\det
(M_{n+1})}{\det (M_{n})}\right), \\
q_{n}^{[N]}& =\frac{\beta }{2}\left[ \frac{\det (G_{n})}{\det (M_{n})}\right]
\mathrm{e}^{\mathrm{i}\theta _{n}}, \\
x_{n}^{[N]}& =\frac{\gamma }{2}an-\frac{\beta ^{2}}{8}s-\frac{2}{a}\ln
_{s}\det (M_{n}),\,\,t=-s,
\end{split}%
\end{equation}
where
\begin{equation*}
\begin{split}
M_{n}& =\left( \frac{\mathrm{e}^{\theta _{m,n}^{\ast }+\theta _{k,n}}}{\eta
_{m}^{\ast }-\eta _{k}}-\frac{\mathrm{e}^{\theta _{m,n}^{\ast }}}{\eta
_{m}^{\ast }-\chi _{k}}-\frac{\mathrm{e}^{\theta _{k,n}}}{\chi _{m}^{\ast
}-\eta _{m}}+\frac{1}{\chi _{m}^{\ast }-\chi _{k}}\right) _{1\leq m,k\leq
N},\,\, \\
G_{n}& =\left( \frac{\mathrm{e}^{\theta _{m,n}^{\ast }+\theta _{k,n}}}{\eta
_{m}^{\ast }-\eta _{k}}\frac{\eta _{m}^{\ast }+c}{\eta _{k}+c}-\frac{\mathrm{%
e}^{\theta _{m,n}^{\ast }}}{\eta _{m}^{\ast }-\chi _{k}}\frac{\eta
_{m}^{\ast }+c}{\chi _{k}+c}-\frac{\mathrm{e}^{\theta _{k,n}}}{\chi
_{m}^{\ast }-\eta _{m}}\frac{\chi _{m}^{\ast }+c}{\eta _{k}+c}+\frac{1}{\chi
_{m}^{\ast }-\chi _{k}}\frac{\chi _{m}^{\ast }+c}{\chi _{k}+c}\right)
_{1\leq m,k\leq N},
\end{split}%
\end{equation*}%
the parameters $\theta _{k,n}$, $\eta _{i}$, $\chi _{i}$ are given in
equations \eqref{theata}.
\end{prop}
Finally, we provide a proof that the above multi-breather solution will converge to the multi-breather solution to the CSP equation obtained in \cite{LFZPhysD}. To this end, we assume $a=b \to 0$ in the continuous limit and notice that $\gamma \to -\gamma$
%and a scaling of $2/a$ for the variable $s$
in compared with the dark soliton solution in \cite{LFZPhysD}. By using the Taylor expansion \eqref{Taylor}, $\theta _{i,n}$ becomes
\begin{equation}\label{breatherdiscon}
    \begin{split}
&   \frac{\mathrm{i}}{2}\xi _{i}s+ n\ln \left( {\frac{\sin (\frac{%
b}{2})\left(\frac{1}{2}\mathrm{i}a\gamma -\xi _{i}\right) +\mathrm{i}\cos (\frac{b}{2})\lambda _{i}}{\sin (\frac{b}{2})\left( \frac{1}{2}\mathrm{i}a{%
\gamma }+\xi _{i}\right) +\mathrm{i}\cos (\frac{b}{2})\lambda _{i}}}\right)
+a_i\\
& \approx \frac{\mathrm{i}}{2}\xi _{i}s-2n \frac{\sin (\frac{b}{2}) \xi_{i}}{\mathrm{i}\cos (\frac{b}{2})\lambda _{i} + \frac{1}{2}\mathrm{i}a\gamma \sin (\frac{b}{2}) }+a_i\\
& = \frac{\mathrm{i}}{2}\xi _{i}s+ \frac{\mathrm{i} \xi_{i}}{\lambda_i} y+a_i
    \end{split}
\end{equation}
by letting $nb=y$ in the continuous limit $b \to 0$. This shows how the multi-breather solution to semi-discrete CSP equation converges to the multi-breather solution of the CSP equation in the continuous limit.
%Therefore $\theta_{i,n}$ agrees with (32) in \cite{LFZPhysD} by noticing the correspondence $\theta_{i,n} \to 2\theta_{i,n}$ and $\gamma \to -\gamma$.

%{\bf {Here I will prove in the continuous limit $a \to 0$, the multi-breather solution will approach to the continuous case we obtained previously}}

\section{Fundamental and high-order rogue wave solution}
\label{section6}
In this section, we will derive the general rogue wave solution for the focusing semi-discrete CSP equation based on the general breather solution obtained in the previous section.  Since the solution vectors involve the square root of a complex
number, it is inconvenient to calculate. To avoid this trouble, we introduce the following transformation:
\begin{equation*}
\lambda _{i}+c=\beta \sinh \left[ \frac{1}{2}(\varphi _{i}^{R}+\mathrm{i}%
\varphi _{i}^{I})\right] ,\,\,(\varphi _{i}^{R},\varphi _{i}^{I})\in \Omega ,
\end{equation*}%
where $\Omega =\{(\varphi ^{R},\varphi ^{I})|0<\varphi ^{I}<\pi ,\,\text{and
}0<\varphi ^{R}<\infty ,\text{or }\varphi ^{R}=0,\text{and }\frac{\pi }{2}%
\leq \varphi ^{I}<\pi \}$, then
\begin{equation*}
\xi _{i}=\beta \cosh \left[ \frac{1}{2}(\varphi _{i}^{R}+\mathrm{i}\varphi
_{i}^{I})\right] ,\,\,\eta _{i}+c=\beta \mathrm{e}^{\frac{1}{2}(\varphi
_{i}^{R}+\mathrm{i}\varphi _{i}^{I})},\,\,\chi _{i}+c=-\beta \mathrm{e}^{-%
\frac{1}{2}(\varphi _{i}^{R}+\mathrm{i}\varphi _{i}^{I})}.
\end{equation*}

Actually, we can obtain the rogue wave solution and high order rogue wave
solutions in this special point. The general procedure to yield these
solutions was proposed in \cite{Guo1,Guo2}. If we solve the linear system %
\eqref{cd-lax} with $(q_{n},\rho _{n},\lambda )=(q_{n}^{[0]},\rho
_{n}^{[0]},-c+\mathrm{i}\beta )$, where $q_{n}^{[0]}$ and $\rho _{n}^{[0]}$
are given in equations \eqref{seed2}, then the quasi-rational solution
vector is obtained. With this solution vector, we could construct the first
order rogue wave solution but fails to obtain
the high order RW solutions. To obtain the general high order rogue wave
solution with a simple way, we must solve the linear system \eqref{cd-lax}
with $(q_{n},\rho _{n},\lambda )=(q_{n}^{[0]},\rho _{n}^{[0]},-c+\mathrm{i}%
\beta \cos (\epsilon ))$, where $\epsilon $ is a small parameter. Denote
\begin{equation}
\lambda _{1}=-c+\mathrm{i}\beta \cos (\epsilon ),\,\,\xi _{1}=\beta \sin
(\epsilon ),\,\,\eta _{1}=\lambda _{1}+\xi _{1}=-c+\mathrm{i}\beta \mathrm{e}%
^{-\mathrm{i}\epsilon },\,\,c=-\frac{1}{2}a\gamma \cot (\frac{b}{2}).
\label{subs1}
\end{equation}

\begin{lem}
\label{lem2} The following parameters can be expanded with $\epsilon$, where
$\epsilon$ is a small parameter
\begin{equation*}
\begin{split}
\mu_1& =\sum_{n=0}^{\infty}\mu_1^{[n]}\epsilon^{2n+1}, \\
\frac{\beta}{\mathrm{i}(\eta_1^*-\eta_1)}&=\frac{1}{\mathrm{e}^{\mathrm{i}%
\epsilon^*}+\mathrm{e}^{-\mathrm{i}\epsilon}}=\sum_{i=0,j=0}^{\infty,%
\infty}F^{[i,j]}\epsilon^{*i}\epsilon^j,
\end{split}%
\end{equation*}
where
\begin{equation*}
\begin{split}
\mu_1^{[n]}&=\frac{\beta\left(-1\right)^n}{(2n+1)!}, \\
F^{[i,j]} &=\frac{1}{i!j!}\frac{\partial^{i+j}}{\partial\epsilon^{*i}%
\partial\epsilon^{j}}\left(\left[\mathrm{e}^{\mathrm{i}\epsilon^*}+\mathrm{e}%
^{-\mathrm{i}\epsilon}\right]^{-1}\right)_{|_{\epsilon^*=0,\epsilon=0}},
\end{split}%
\end{equation*}
\end{lem}
With the aid of above lemma \ref{lem2}, we obtain the following expansion
\begin{equation*}
\begin{split}
Z_{1,n}& \equiv \frac{\mathrm{i}\beta}{4} \sin\left(\epsilon\right)s+\frac{n%
}{2}\ln\left({\frac {\beta \sin \left( \frac{3}{2} b-\epsilon \right) +\beta
\sin \left( \frac{1}{2} b+\epsilon \right) -2 \mathrm{i}\cos \left( \frac{1}{%
2} b \right) a{\gamma}}{ \beta \sin \left( \frac{3}{2} b+\epsilon \right)
+\beta \sin \left( \frac{1}{2} b -\epsilon \right) -2 \mathrm{i}\cos \left(
\frac{1}{2} b \right) a{\gamma}}} \right) -\frac{\mathrm{i}\epsilon}{2}
+\sum_{i=1}^{\infty} (e_i+\mathrm{i}f_i)\epsilon^{2i-1} , \\
&=\mathrm{i}\epsilon\sum_{k=0}^{\infty}Z_{1,n}^{[2k+1]}\epsilon^{2k},
\end{split}%
\end{equation*}
where
\begin{equation*}
Z_{1,n}^{[2k+1]}=\frac{\mathrm{d}^{2k+1}}{\mathrm{d}\epsilon^{2k+1}}%
Z_{1,n}|_{\epsilon=0}.
\end{equation*}

Furthermore we have
\begin{equation*}
\mathrm{e}^{Z_{1,n}}=\sum_{i=0}^{\infty}S_i(\mathbf{Z}_{1,n})\epsilon^i,\,\,
\mathbf{Z}_{1,n}=\left(Z_{1,n}^{[1]},Z_{1,n}^{[2]},\cdots\right),\,\,
Z_{1,n}^{[2k]}=0,\,\, k\geq 1 \\
\end{equation*}
the explicit expression of these polynomials can be are given by the
elementary Schur polynomials
\begin{equation*}
\begin{split}
S_0(\mathbf{Z}_{1,n})=&1,\,\, S_1(\mathbf{Z}_{1,n})=Z_{1,n}^{[1]},\,\, S_2(%
\mathbf{Z}_{1,n})=Z_{1,n}^{[2]}+\frac{(Z_{1,n}^{[1]})^2}{2},\,\, S_3(\mathbf{%
Z}_{1,n})=Z_{1,n}^{[3]}+Z_{1,n}^{[1]}Z_{1,n}^{[2]}+\frac{(Z_{1,n}^{[1]})^3}{6%
},\cdots \\
S_i(\mathbf{Z}_{1,n})=&\sum_{l_1+2l_2+\cdots+kl_k=i}\frac{%
(Z_{1,n}^{[1]})^{l_1}(Z_{1,n}^{[2]})^{l_2}\cdots (Z_{1,n}^{[k]})^{l_k}}{%
l_1!l_2!\cdots l_k!}.
\end{split}%
\end{equation*}
Since $KE_{1,n}(\epsilon)$ satisfies the Lax equation \eqref{cd-lax}, then $%
KE_{1,n}(-\epsilon)$ also satisfies the Lax equation \eqref{cd-lax}. To
obtain the general high order rogue wave solution, we choose the general
special solution
\begin{equation*}
|y_{1,n}\rangle=\frac{K}{2\epsilon}\left[E_{1,n}(\epsilon)-E_{1,n}(-\epsilon)%
\right]\equiv K%
\begin{bmatrix}
\varphi_{1,n} \\[8pt]
\beta\psi_{1,n} \\
\end{bmatrix}%
,
\end{equation*}
where
\begin{equation*}
E_{1,n}=%
\begin{bmatrix}
\mathrm{e}^{Z_{1,n}} \\
{\displaystyle \frac{\beta\mathrm{e}^{Z_{1,n}}}{\eta_1+c}} \\
\end{bmatrix},
\end{equation*}
Finally, we have
\begin{equation}  \label{expansion1}
\begin{split}
\frac{\beta\langle y_{1,n}|y_{1,n}\rangle}{2\mathrm{i}(\lambda_1^*-\lambda_1)%
} & =\frac{\beta}{4\mathrm{i}}\left[\frac{\mathrm{e}^{Z_{1,n}^*+Z_{1,n}}}{%
\eta_{1}^*-\eta_{1}}-\frac{\mathrm{e}^{Z_{1,n}^*-Z_{1,n}}}{%
\eta_{1}^*-\chi_{1}}-\frac{\mathrm{e}^{-Z_{1,n}^*+Z_{1,n}}}{%
\chi_{1}^*-\eta_{1}}+\frac{\mathrm{e}^{-Z_{1,n}^*-Z_{1,n}}}{%
\chi_{1}^*-\chi_{1}}\right] \\
&=\sum_{m=1,k=1}^{\infty,\infty}M_n^{[m,k]} {\epsilon}^{*2(m-1)}%
\epsilon^{2(k-1)}
\end{split}%
\end{equation}
and
\begin{equation}  \label{expansion2}
\begin{split}
&\frac{\mathrm{i}\beta\langle y_{1,n}|y_{1,n}\rangle}{2(\lambda_1^*-%
\lambda_1)}+\mathrm{i}\beta\varphi_{1,n}\psi_{1,n} \\
=&\frac{\mathrm{i}\beta}{4}\left[\frac{\mathrm{e}^{Z_{1,n}^*+Z_{1,n}}}{%
\eta_{1}^*-\eta_{1}}\frac{\eta_{1}^*+c}{\eta_{1}+c}-\frac{\mathrm{e}%
^{Z_{1,n}^*-Z_{1,n}}}{\eta_{1}^*-\chi_{1}}\frac{\eta_{1}^*+c}{\chi_{1}+c}-%
\frac{\mathrm{e}^{-Z_{1,n}^*+Z_{1,n}}}{\chi_{1}^*-\eta_{1}}\frac{\chi_{1}^*+c%
}{\eta_{1}+c}+\frac{\mathrm{e}^{-Z_{1,n}^*-Z_{1,n}}}{\chi_{1}^*-\chi_{1}}%
\frac{\chi_{1}^*+c}{\chi_{1}+c}\right] \\
=&\sum_{m=1,k=1}^{\infty,\infty}G_n^{[m,k]} {\epsilon}^{*2(m-1)}%
\epsilon^{2(k-1)}
\end{split}%
\end{equation}
where $\chi_1=\eta_1(-\epsilon),$
\begin{equation}  \label{rogue}
\begin{split}
M_n^{[m,k]}&=\sum_{i=0}^{2m-1}\sum_{j=0}^{2k-1}F^{[i,j]}S_{2k-i-1}(\mathbf{Z}%
_{1,n})S_{2m-j-1}(\mathbf{Z}_{1,n}^*), \\
G_n^{[m,k]}&=\sum_{i=0}^{2m-1}\sum_{j=0}^{2k-1}F^{[i,j]}S_{2k-i-1}(\mathbf{Z}%
_{1,n}+\mathbf{\varepsilon})S_{2m-j-1}(\mathbf{Z}_{1,n}^*+\mathbf{\varepsilon%
}),
\end{split}%
\end{equation}
and $\mathbf{\varepsilon}=(1,0,0,\cdots).$

Based on the expansion equations \eqref{expansion1}-\eqref{expansion2}, and
formulas \eqref{gBT}-\eqref{linalglem}, we can obtain the general rogue wave
solutions:
\begin{prop}
\label{prop4} The general high order rogue wave solution for semi-discrete
CSP equation \eqref{sem-csp} can be represented as
\begin{equation}  \label{gene-rogue1}
\begin{split}
\rho_n^{[N]}&=\frac{\gamma}{2}-\frac{2}{a}\ln_s\left(\frac{\det(M_{n+1})}{%
\det(M_{n})}\right)>0, \\
q_n^{[N]}&=\frac{\beta}{2}\left[\frac{\det(G_n)}{\det(M_n)}\right]\mathrm{e}%
^{\mathrm{i}\theta_n}, \\
x_n^{[N]}&=\frac{\gamma}{2} a n-\frac{\beta^2}{8}s-\frac{2}{a}%
\ln_{t}\det(M_n),\,\, t=-s,
\end{split}%
\end{equation}
where
\begin{equation*}
M_n=\left(M_n^{[m,k]}\right)_{1\leq m,k\leq N},\,\,
G_n=\left(G_n^{[m,k]}\right)_{1\leq m,k\leq N},
\end{equation*}
the expressions $M_n^{[m,k]}$ and $G_n^{[m,k]}$ are given in equations %
\eqref{rogue}.
\end{prop}

Specially, the first order rogue wave solution can be written explicitly
through formula \eqref{gene-rogue1}
\begin{equation*}
\begin{split}
\rho_n^{[1]}=&\frac{\gamma}{2}-\frac{2}{a}\ln_s\left(\frac{\frac{1}{4}%
+(Z_{n+1,R}^{[1]})^2+(Z_{n+1,I}^{[1]}+\frac{1}{2})^2}{\frac{1}{4}%
+(Z_{n,R}^{[1]})^2+ (Z_{n,I}^{[1]}+\frac{1}{2})^2}\right), \\
q_n^{[1]}=&\frac{\beta}{2}\left[1-\frac{1-2\mathrm{i}Z_{n,R}^{[1]}}{\frac{1}{%
4}+(Z_{n,R}^{[1]})^2+(Z_{n,I}^{[1]}+\frac{1}{2})^2}\right]\mathrm{e}^{%
\mathrm{i}\theta}, \\
x_n^{[1]}=&\frac{\gamma}{2} a n-\frac{\beta^2}{8}s-\frac{2}{a}\ln_{s}\left(%
\frac{1}{4}+(Z_{n,R}^{[1]})^2+ (Z_{n,I}^{[1]}+\frac{1}{2})^2\right),\,\,
t=-s,
\end{split}%
\end{equation*}
where
\begin{equation}  \label{expx1}
\begin{split}
Z_{n,R}^{[1]}= & \frac{ 4 {\beta}^{2}\sin^{3}(\frac{b}{2})\cos(\frac{ b}{2})
n}{{a}^{2}{\gamma}^{2}+2 {\ \beta}^{2}\sin^{2}(b)}, \\
Z_{n,I}^{[1]}= & \beta \left(\frac{2a\gamma\sin^{2}(\frac{b}{2}) n}{{a}^{2}{%
\gamma}^{2}+2 {\ \beta}^{2}\sin^{2}(b)}+\frac{s}{4}\right)-\frac{1}{2}.
\end{split}%
\end{equation}
Moreover, the general second order rogue wave solution can be represented by the formula \eqref{gene-rogue1}:
\begin{equation*}
\begin{split}
\rho_n^{[2]}=&\frac{\gamma}{2}-\frac{2}{a}\ln_s\left(\frac{F_{1,n+1}+\mathrm{%
i}F_{2,n+1}}{F_{1,n}+\mathrm{i}F_{2,n}}\right)>0, \\
q_n^{[2]} =&\frac{\beta}{2}\left[1+\frac{G_n}{F_{1,n}+\mathrm{i}F_{2,n}}%
\right]\mathrm{e}^{\mathrm{i}\theta_n}, \\
x_n^{[2]}=&\frac{\gamma}{2} a n-\frac{\beta^2}{8}s-\frac{2}{a}%
\ln_{s}\left(F_{1,n}+\mathrm{i}F_{2,n}\right),\,\, t=-s,
\end{split}%
\end{equation*}
where
\begin{equation*}
\begin{split}
F_{1,n}=&\left( -{\frac {1}{72}} \overline{Z_n^{[1]}}-\frac{1}{12}\overline{%
Z_n^{[3]}}+\frac{1}{36} {\overline{Z_n^{[1]}}}^{3} \right) ({Z_n^{[1]}}%
)^{3}+ \left(\frac{1}{8} {\overline{Z_n^{[1]}}}^{2}-{\frac {1}{64}} \right) (%
{Z_n^{[1]}})^{2}+ \left(-{\frac {1}{72}} {\overline{Z_n^{[1]}}}^{3}+\frac{1}{%
24} \overline{Z_n^{[3]}}+{\frac {29}{288}} \overline{Z_n^{[1]}} \right)
Z_n^{[1]} \\
&+\frac{1}{4} Z_n^{[3]}\overline{Z_n^{[3]}}-\frac{1}{12} Z_n^{[3]}{\overline{%
Z_n^{[1]}}}^{3}- {\frac {1}{64}}{\overline{Z_n^{[1]}}}^{2}+\frac{1}{24}
\overline{Z_n^{[1]}}Z_n^{[3]}+{\frac {1}{64}},
\end{split}%
\end{equation*}
\begin{equation*}
F_{2,n}=\frac{1}{24} {Z_n^{[1]}}^{2}{\overline{Z_n^{[1]}}}^{3}+\frac{1}{8} {%
\overline{Z_n^{[1]}}}^{2}Z_n^{[3]}-\frac{1}{24}{\overline{Z_n^{[1]}}}^{2}{%
Z_n^{[1]}}^{3}-\frac{1}{8} {Z_n^{[1]}}^{2}\overline{Z_n^{[3]}}+\frac{1}{12}
Z_n^{[1]}{\overline{Z_n^{[1]}}} ^{2}-\frac{1}{12} {Z_n^{[1]}}^{2}\overline{%
Z_n^{[1]}}+\frac{1}{32} \overline{Z_n^{[1]}}-\frac{1}{32} Z_n^{[1]},
\end{equation*}
\begin{equation*}
\begin{split}
G_n=&\frac{1}{12} \mathrm{i}{\overline{Z_n^{[1]}}}^{2}{Z_n^{[1]}}^{3}+
\left( \frac{1}{12} \mathrm{i}\overline{Z_n^{[1]}}-\frac{1}{4} {\overline{%
Z_n^{[1]}}}^{2}-\frac{1}{4} \mathrm{i}\overline{Z_n^{[3]}}+\frac{1}{12}%
\mathrm{i}{\overline{Z_n^{[1]}}}^{3} \right) {Z_n^{[1]}}^{2}+ \left( \frac{1%
}{2} \overline{Z_n^{[3]}}- \frac{1}{6}{\overline{Z_n^{[1]}}}^{3}-\frac{1}{6}%
\mathrm{i}{\overline{Z_n^{[1]}}}^{2}-\frac{1}{6} \overline{Z_n^{[1]}}
\right) Z_n^{[1]} \\
&-\frac{1}{4} \mathrm{i}{\overline{Z_n^{[1]}}}^{2}Z_n^{[3]}+\frac{1}{4}
\mathrm{i}\overline{Z_n^{[3]}}-\frac{1}{12}\mathrm{i}\overline{Z_n^{[1]}}-%
\frac{1}{12}\mathrm{i}{\overline{Z_n^{[1]}}}^{3},
\end{split}%
\end{equation*}
$Z_n^{[1]}=Z_{n,R}^{[1]}+\mathrm{i}Z_{n,I}^{[1]}$, the symbol overbar
denotes the complex conjugation and
\begin{equation*}
Z_{n}^{[3]}=\frac{\frac{2}{3}\mathrm{i} \beta\sin^{2}(\frac{1}{2}b)
\left(2a\beta\gamma\sin(b)+\mathrm{i}{a}^{2}\gamma^{2}+8 \mathrm{i}{\ \beta}%
^{2}\sin^{2}(\frac{1}{2}b) \right)}{2 \beta {a}^{2}{\gamma}^{2} \sin(b)-{%
\beta}^{3} \sin^3(b) -\mathrm{i}{a}^{3}{\gamma}^{3}+3 \mathrm{i}a\beta^{2}
\gamma\sin^{2}(b) }\frac{n}{2}-\frac{\mathrm{i}\beta s}{24}+(e_1+f_1\mathrm{i%
}).
\end{equation*}

\begin{figure}[tbh]
\centering
\subfigure[K-M breather]{\includegraphics[height=50mm,width=65mm]{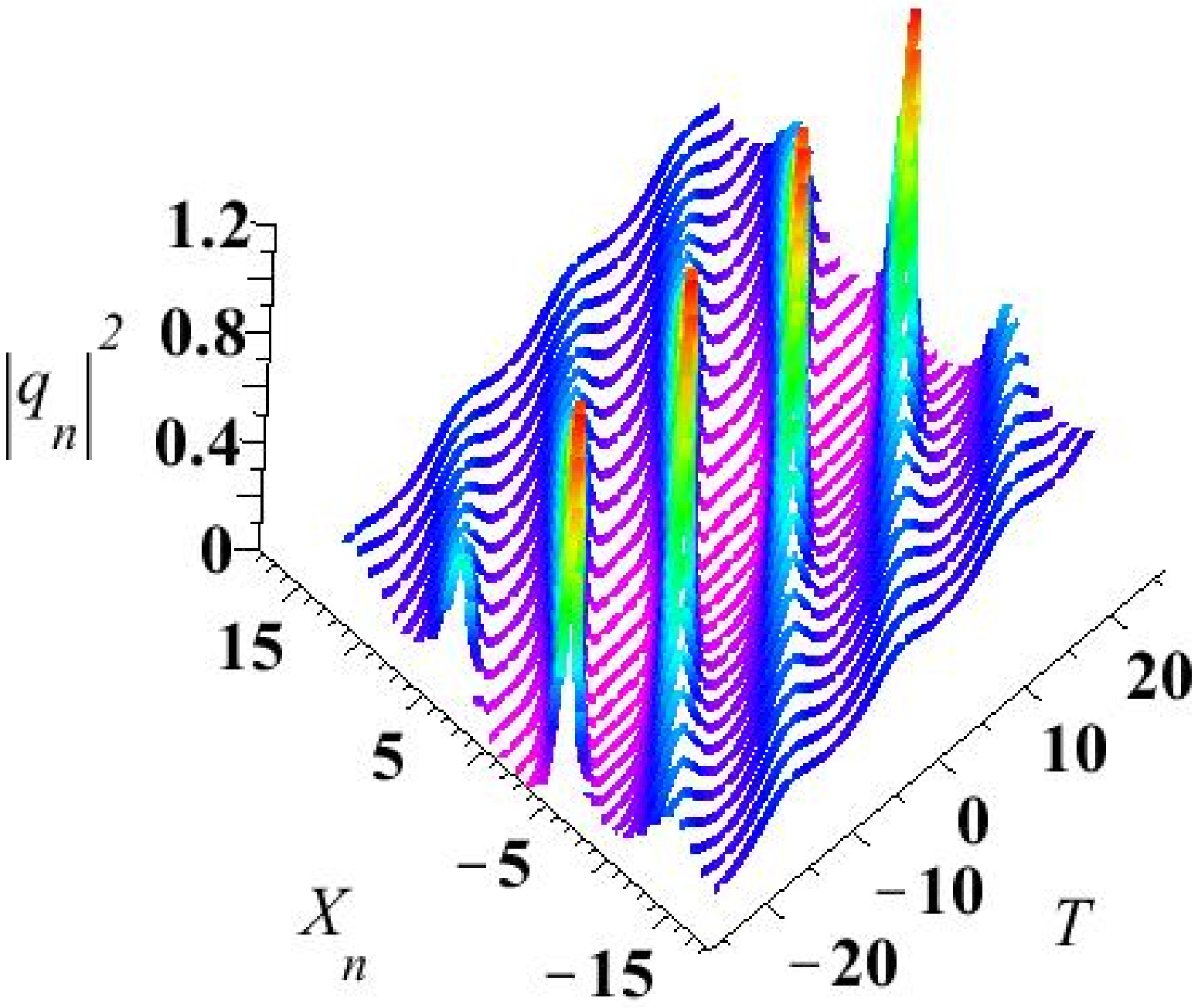}}
\hfil
\subfigure[Fundamental RW]{\includegraphics[height=50mm,width=65mm]{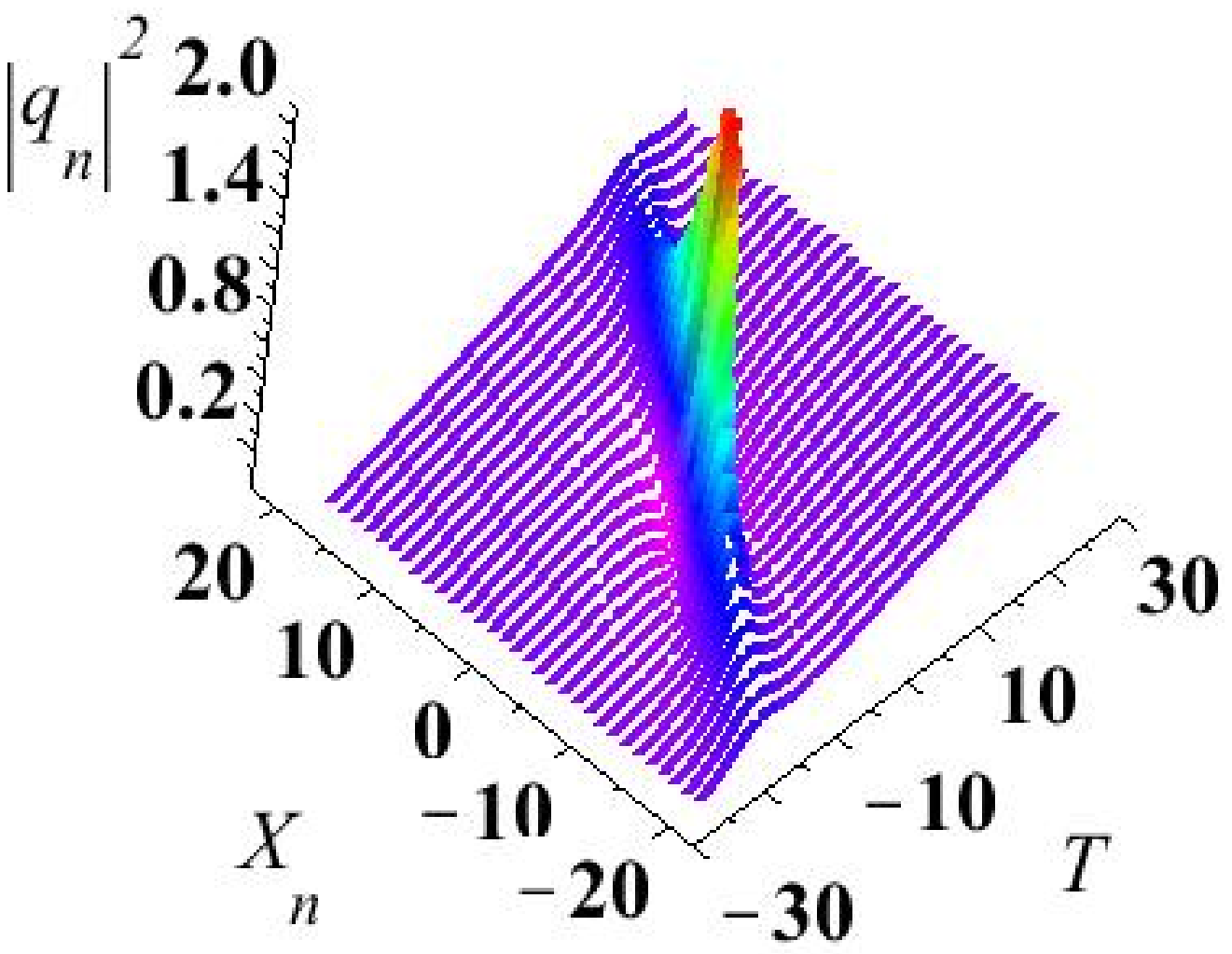}}
\caption{(color online): Breather and Rogue waves}
\label{figure2}
\end{figure}

\begin{figure}[tbh]
\centering
\subfigure[Second order RW]{\includegraphics[height=50mm,width=65mm]{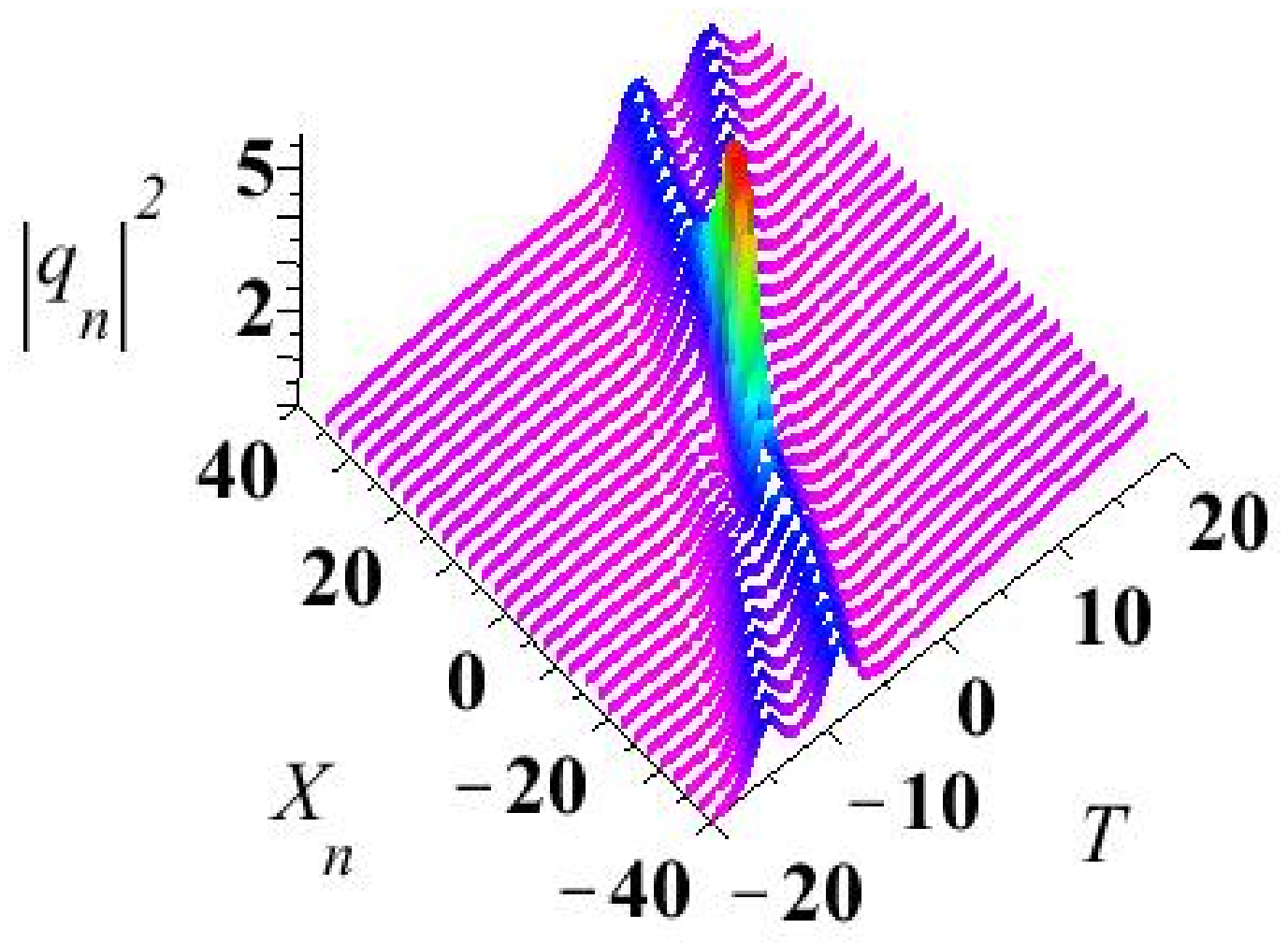}}
\hfil
\subfigure[Second order RW]{\includegraphics[height=50mm,width=65mm]{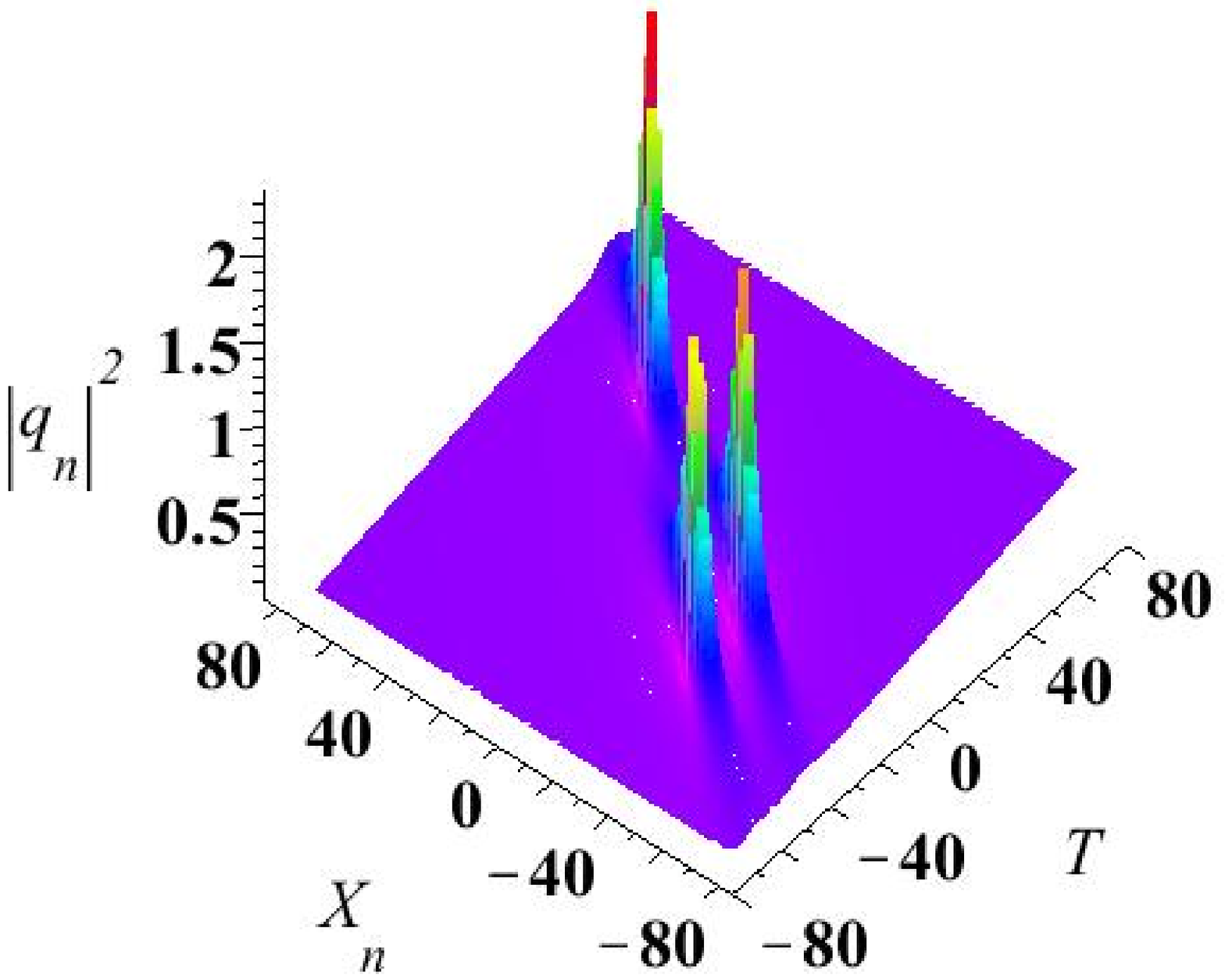}}
\caption{(color online): Second order rogue waves with different dynamics}
\label{figure3}
\end{figure}

To illustrate the dynamics of rogue waves, we firstly show the
fundamental rogue wave in Fig. \ref{figure2}(b)  with the parameters $a=2$, $b=%
\frac{\pi}{2}$, $\beta=1$, $\gamma=\frac{5}{4}$. For the second order rogue
waves, we firstly choose the parameters $a=2$, $b=\frac{\pi}{2}$, $\beta=1$,
$\gamma=\frac{5}{2}$, $e_1=f_1=0$, then the standard second order RWs is
shown in Fig. \ref{figure3} (a). To exhibit the other dynamics for the second
order RWs, we choose the parameters $a=2$, $b=\frac{\pi}{2}$, $\beta=1$, $%
\gamma=\frac{3}{2}$, $e_1=10$, $f_1=0$. It is seen that the temporal-spatial
distribution exhibit the triangle shape as shown in Fig. \ref{figure3} (b).
%Here we
%point out the figure \ref{figure3}(b) looks like the figure of continuing
%function since the lattices for it are so dense.

We remark here that since the higher order rogue wave solution is obtained from multi-breather solution to the semi-discrete CSP equation which converges to its counterpart in the continuous CSP equation, thus, the high order rogue wave solution for the semi-discrete CSP equation should converge to the one for CSP equation in the continuous limit.

\section{Conclusions and discussions}
\label{section7} In the present paper, we firstly drive the generalized Darbourx transformation (gDT) for the semi-discrete CSP equation \eqref{sem-csp} with the aid of discrete
hodograph transformation. %It is well known that the DT is a powerful method
%to construct the solitonic solutions for the integrable system. Comparing
%with the other method, it is more convenient to construct the solutions
%with the uniform formulas.
Based on formulas derived from the gDT, we then construct the multi-bright soliton solution for the focusing CSP equation with zero boundary condition. For the nonzero boundary conditions, we construct the multi-dark soliton solution for the defocusing case, the multi-breather solution and high-order rogue wave solution for the focusing case.  We
require the condition $\rho_n>0$ to keep the non-singularity and monodromy.
Otherwise, if $\rho_n$ does not keep the positive definitive property, then
the cusp or loop soliton  would appear. All above solutions are shown to converge to their counterparts for the original CSP equation in the continuous limit.

It is noticed that a robust inverse scattering transform method has been proposed for the NLS equation by appropriately setting up and solving the Riemann-Hilbert problem\cite{BP,BLM}. It is imperative to study the inverse scattering transform and Riemann-Hilbert problem for both the original and semi-discrete CSP equation. Although the multi-bright soliton solutions have been constructed by Hirota's bilinear method in determinant form \cite{FMO-PJMI} and in pfaffian form \cite{FMOmultiSP}, it would be interesting to drive other types of solutions such as dark-soliton, breather and rogue wave solutions for the semi-discrete CSP equation.

%Comparing the results with CCD
%equations, we can find that the solitonic solutions are maintained under the process of discretization.

In the last, we should point out the following coupled semi-discrete CSP equation
\begin{equation} \label{sem-csp21}
\begin{split}
& \frac{d}{dt}(q_{1,n+1}-q_{1, n})=\frac{1}{2}(x_{n+1}-x_{n})(q_{1, n+1}+q_{1, n}), \\
& \frac{d}{dt}(q_{2,n+1}-q_{2, n})=\frac{1}{2}(x_{n+1}-x_{n})(q_{2, n+1}+q_{2, n}), \\
& \frac{d}{dt}(x_{n+1}-x_{n})+\frac{1}{2}
\sum_{j=1}^2 \sigma_j \left(|q_{j,n+1}|^{2}-|q_{j,n}|^{2}\right)=0\,,
\end{split}%
\end{equation}
which has been shown to be integrable recently \cite{FMOmultiSP}. Beside the multi-bright soliton solution implied in \cite{FMOmultiSP}, how about its general initial value problem and other types of soliton solutions?

The method provided in this paper is also useful
to the coupled semi-discrete CSP equation. We expect to obtain and report  the results in the near future. As the last conment,
The obtained semi-discrete equations can be served as superior numerical schemes: the so-called self-adaptive moving mesh schemes for the CSP and coupled CSP equations.

\section*{Acknowledgments}
B.-F. F. acknowledges the partial support by NSF under Grant No. DMS-1715991, and National Natural
Science Foundation of China under Grant No.11728103.
The work of L.M.L. is supported by National Natural Science Foundation of China (Contact Nos. 11771151),
Guangdong Natural Science Foundation (Contact No. 2017A030313008), Guangzhou Science and Technology
Program(No. 201707010040). The work of Z.N. Z is partially supported by National Natural Science Foundation of China (No. 11671255) and by the Ministry of Economy and Competitiveness of Spain under contract
MTM2016-80276-P (AEI/FEDER, EU).


\begin{thebibliography}{99}
\bibitem{2018Nobelprize} S., Donna, M., Gerard, Compression of amplified chirped optical pulses, \textit{Optics Commun.}, \textbf{56} 219?221 (1985).

\bibitem{Agrawalbook} G.~P.~Agrawal, \textit{Nonlinear Fiber Optics},
(Academic Press, New York, 1995).

\bibitem{HasegawaKodama} A.~Hasegawa and Y.~Kodama, \textit{Solitons in
Optical Communications}, (Clarendon, Oxford, 1995).

\bibitem{AgrawalKivsharbook} Y~S.~Kivshar, G.~P.~Agrawal, \textit{optical
Solitons: From Fibers to Photonic Crystals}, (Academic Press, San Diego,
2003).

\bibitem{HasegawaTappert1} A.~Hasegawa and F.~Tappert,
Transmission of stationary nonlinear optical pulses in dispersive dielectric fibers I. Anomolous dispersion,
\textit{Appl. Phys. Lett.} \textbf{23} 142 (1973).

\bibitem{HasegawaTappert2} A.~Hasegawa and F. Tappert,
Transmission ofstationary nonlinear optical pulses in dispersive dielectric fibers II. Normal dispersion,
\textit{Appl. Phys. Lett.} \textbf{23}  171 (1973).

\bibitem{Ablowitzbook} M.~J.~Ablowitz, P.~A.~ Clarkson, \textit{Solitons,
Nonlinear Evolution Equations and Inverse Scattering} (\textit{London
Mathematical Society Lecure Notes Series} \textbf{149}), (Cambridge Univ.
Press, Cambridge, 1991).

\bibitem{APT} M.~J.~Ablowitz, B.~Prinari, and A.~D.~Trubatch, \textit{
Discrete and continuous nonlinear Schr\"{o}dinger systems}, (Cambridge Univ.
Press, Cambridge, 2004).


\bibitem{BECReview} F.~Dalfovo, S.~Giorgini and L.~P.~ Stringari,
Theory of Bose-Einstein condensation in trapped gases,
\textit{Rev. Mod. Phys.} \textbf{71}  463--512 (1999).

\bibitem{Benney1967} D.~J.~Benney and A.~C.~Newell,
The propagation ofnonlinear wave envelopes,
\textit{Stud. Appl. Math.} \textbf{46}  133--139 (1967).

\bibitem{ZakharovPlasma} V.~E.~Zakharov, Collapse of Langumuir waves,
\textit{Sov. Phys. JETP} \textbf{35}  908--914 (1972).

\bibitem{ZakharovShabat} V.~E.~Zakharov and A.~B.~Shabat,
Exact theory of two-dimensional self-focusing and one-dimensional self-modulations of waves in nonlinear media,
%\it{Zh. Eksp. Teor. Fiz.} \textbf{61}, 118 (1971) [
\textit{Sov. Phys. JETP} \textbf{34}  62--69 (1972).

\bibitem{ZakharovShabat2} V.~E.~Zakharov and A.~B.~Shabat,
Interaction betweem solitons in a stable medium,
\textit{Sov. Phys. JETP} \textbf{37} 823--828 (1973).

%\bibitem{HasegawaTappert} A.~Hasegawa and F.~Tappert, Transmission of %stationary nonlinear
%optical pulses in dispersive dielectric fibers. II. Normal
%dispersion, \textit{Appl. Phys. Lett.}  \textbf{23} (1973), 171.

\bibitem{Krokeldark1} D. Krokel, N.~J. Halas, G.~Giuliani and D.~Grischkowsky,
Dark-pulse propagation in optical fibers, \textit{Phys. Rev. Lett.} \textbf{60}  29 (1988).

\bibitem{Weinerdark1} A.~M.~Weiner, J.~P.~Heritage, R.~J.~Hawkins, R.~N.~Thurston, E.~M.~Kirschner, D.~E.~Leaird
 and W. J. Tomlinson, Experimental observation of the fundamental dark soliton in optical fibers,
\textit{Phys. Rev. Lett.} \textbf{61}  2445 (1988).

\bibitem{Opticalroguewave} O. R. Solli, C. Ropers, P. Koonath, B. Jalali,
Optical rogue waves, \textit{Nature}, \textbf{450} 1054--1057 (2007).

\bibitem{Peregrinesoliton} B. Kibler, J. Fatome, C. Finot, G. Millot, F.
Dias, G. Genty, N. Akhmediev, J. M. Dudley, The Peregrine soliton in
nonlinear fibre optics, \textit{Nature Physics}, \textbf{6} 790--795 (2010).

\bibitem{AblowitzLadik} M. J. Ablowitz and J. F. Ladik,  Nonlinear
differential-difference equations, \textit{J. Math. Phys.} \textbf{16}
598--603 (1975).

\bibitem{AL2} M. J. Ablowitz and J. F. Ladik, Nonlinear
differential-difference equations and Fourier analysis, \textit{J. Math. Phys.}
\textbf{17}  1011-1018 (1976).

\bibitem{TsujimotoBookchapter} S. Tsujimoto,  Chap. 1 in \textit{Applied Integrable Systems}, Ed. Y. Nakamura, (Shokabo, Tokyo, 2000) [in Japanese].

\bibitem{Narita1990} K. Narita,
Soliton solution for discrete Hirota equation
\textit{J. Phys. Soc. Jpn.} \textbf{59}  3528--3530 (1990).

\bibitem{OhtaMaruno} K. Maruno and Y. Ohta, Casorati Determinant Form of
dark soliton solutions of the discrete nonlinear Schr\"{o}dinger equation,
\textit{J. Phys. Soc. Jpn.} \textbf{75} 054002 (2006) .


\bibitem{Konotop92} V.~E.~Vekslerchik and V.~V.~Konotop, Discrete nonlinear Schrodinger equation under
non-vanishing boundary conditions, \textit{Inv. Prob.} \textbf{8}  889-909 (1992).

\bibitem{ABB07} M.~J.~Ablowitz, G.~Biondini and B.~Prinari, Inverse scattering transform for
the integrable discrete nonlinear Schr\"odinger equation with non-vanishing boundary
conditions, \textit{Inv. Prob.} \textbf{23}  1711--1758 (2007).

\bibitem{Mee15} C.~Van Der Mee, Inverse scattering transform for
the discrete focusing nonlinear Schr\"odinger equation with non-vanishing boundary
conditions,
\textit{J. Nonlinear Math. Phys.} \textbf{22}  233--264 (2015).

\bibitem{BPrinari} B.~Prinari, F.~Vitale, Inverse scattering transform for the focusing
Ablowitz-Ladik system with nonzero boundary conditions,
\textit{Stud. Appl. Math.} \textbf{137}  28--52 (2015).

\bibitem{BPrinari2} B.~Prinari, Discrete solitons of the focusing Ablowitz-Ladik equation with nonzero
boundary conditions via inverse scattering,
\textit{J. Math. Phys.} \textbf{57}  083510 (2016).


%\bibitem{BPrinari2} B.~Prinari, Discrete solitons of the focusing %Ablowitz-Ladik equation with nonzero
%boundary conditions via inverse scattering, J. Math. Phys. (2016)

\bibitem{Akhmediev1} A Ankiewicz, N Akhmediev,  JM Soto-Crespo, Discrete rogue waves of the Ablowitz-Ladik and Hirota equations,
\textit{Phys. Rev. E} \textbf{82}  026602 (2010).

\bibitem{Akhmediev2} A Ankiewicz, N Devine, M Unal, A Chowdury and N Akhmediev, Rogue waves and other solutions of single and coupled Ablowitz-Ladik and
nonlinear Schr\"odinger equations,
\textit{J. of Optics} \textbf{15} 064008 (2013).

\bibitem{OhtaYangAL} Y Ohta and J Yang, General rogue waves in the focusing and defocusing Ablowitz-Ladik equations, \textit{J. Phys. A:Math. Theor.}
\textbf{47}  255201 (2014).

\bibitem{DoliwaAL}  A. Doliwa, P. M. Santini,
Integrable dynamics of a discrete curve and the Ablowitz-Ladik hierarchy,
\textit{J. Math. Phys.} \textbf{36}  1259--1273 (1995).

\bibitem{hirotaeq} R. Hirota, Exact envelope-soliton solutions of a nonlinear wave equation, \emph{J. Math. Phys.},
\textbf{14}  805?809 (1973).

\bibitem{sasasatsuma} N. Sasa, J. Satsuma, New-type soliton solution for a higher-order nonlinear
Schr\"odinger equation, \textit{J. Phys. Soc. Jpn.}, \textbf{60}  409?417 (1991).

\bibitem{Rothenberg} J. E. Rothenberg, Space-time focusing: breakdown of the
slowly varying envelope approximation in the self-focusing of femtosecond
pulses, \textit{Opt. Lett.}, \textbf{17}  1340--1342 (1992).

\bibitem{Skobelev} S. A. Skobelev, D. V. Kartashov, A. V. Kim,
Few-optical-cycle solitons and pulse self-compression in a Kerr medium,
\textit{Phys. Rev. Lett.}, \textbf{99}  203902 (2007).

\bibitem{Kim} A. V. Kim, S. A. Skobelev, D. Anderson, T. Hansson, M. Lisak,
Extreme nonlinear optics in a Kerr medium: Exact soliton solutions for a few cycles,
\textit{Phys. Rev. A}, \textbf{77} 043823 (2008).

\bibitem{Amir} S. Amiranashvili, A.~G. Vladimirov, U. Bandelow,
Solitary-wave solutions for few-cycle optical pulses,
\textit{Phys. Rev. A}, \textbf{77} 063821 (2008).

\bibitem{Amir2} S. Amiranashvili,  U. Bandelow, N. Akhmediev
Few-cycle optical solitary waves in nonlinear dispersive media,
\textit{Phys. Rev. A}, \textbf{87} 013805 (2013).

\bibitem{SPE_Org} T. Sch\"afer, C. E. Wayne, Propagation of ultrashort
optical pulses in cubic nonlinear media, \textit{Physica D}, \textbf{196} 90--105 (2004).

\bibitem{Sakovich} A. Sakovich, S. Sakovich, The short pulse equation is
integrable, \textit{J. Phys. Soc. Jpn.}, \textbf{74} 239--241 (2005).

\bibitem{Matsuno_SPEreview} Y. Matsuno, Periodic solutions of the short
pulse model equation, \textit{J. Math. Phys.}, \textbf{49}  (2008).

\bibitem{SPWB} Y. Liu, D. Pelinovsky, A. Sakovich, Wave breaking in the short-pulse equation, \textit{Dynam. Part. Differ. Eq.}, \textbf{6}
291--310 (2009).


\bibitem{d-short} B.-F. Feng, K. Maruno and Y. Ohta, Integrable
discretizations of the short pulse equation, \textit{J. Phys. A}, \textbf{43}
085203 (2010).

\bibitem{SPE_discrete2} B.-F. Feng , J. Inoguchi, K. Kajiwara, K. Maruno, Y. Ohta, Discrete integrable systems and hodograph transformations arising from motions of discrete plane curves, \textit{J. Phys. A}, \textbf{44} 395201 (2011).

\bibitem{Feng_ComplexSPE} B.-F. Feng, Complex short pulse and coupled
complex short pulse equations, \textit{Physica D}, \textbf{297} 62--75 (2015).

\bibitem{FenglingzhuPRE} B.-F. Feng, L. Ling and Z. Zhu, A defocusing
complex short pulse equation and its multi-dark soliton solution by Darboux
transformation, \textit{Phys. Rev. E} \textbf{93}  052227 (2016).

%\bibitem{LingFeng2} B.-F. Feng, L. Ling, Z. Zhu,
%A defocusing complex short pulse equation and its multi-dark soliton solution by Darboux transformation,
%Phys. Rev. E, 93 052227 (2016).

\bibitem{FengShen_ComplexSPE} S. Shen, B.-F. Feng and Y. Ohta, From the real and complex coupled dispersionless equations to the real and complex short
pulse equations, \textit{Stud. Appl. Math.}, \textbf{136}  64--88 (2016).

\bibitem{LFZPhysD} L. Ling, B.-F. Feng and Z. Zhu, Multi-soliton,
multi-breather and higher order rogue wave solutions to the complex short
pulse equation, \textit{Physica D}, \textbf{327} 13--29 (2016).

\bibitem{FMO_ComplexSPE} B.-F. Feng, K. Maruno and Y. Ohta, Geometric
formulation and multi-dark soliton solution to the defocusing complex short
pulse equation, \textit{Stud. Appl. Math.}, \textbf{138} 343--367 (2016).

\bibitem{Optik} B.-Q. Li, Y.-L. Ma, Periodic solutions and solitons to two complex short pulse (CSP) equations in optical fiber, \textit{Optik}, \textbf{144} 149--155 (2017).


\bibitem{Jxu1} J. Xu, Long-time asymptotics  for the  short pulse equation, \textit{J. Diff. Eqn.}, \textbf{265} 3494--3542 (2018).

\bibitem{Jxu2} J. Xu, E. Fan, Long-time asymptotic behavior for the complex short pulse equation, arXiv:1712.07815, 2017.

\bibitem{NolsG} M. J. Ablowitz, Bao-Feng Feng, X.-D. Luo, Z. H.
Musslimani, Reverse space-time Sine/Sinh-Gordon equations with nonzero boundary conditions, \textit{Stud. Appl. Math.},
\textbf{141} 267~307 (2018).

\bibitem{AB} A.M. Kamchatnov, \textit{Nonlinear Periodic Waves and Their
Modulations}, (World Sci. Press, Hong Kong 2000)

\bibitem{Disbook}
J. Hietarinta, N. Joshi, F.W. Nihoff \textit{Discrete Systems and Integrability}, (Cambridge University Press, 2016).

\bibitem{hirota} R. Hirota, {Nonlinear Partial Difference Equations. I. A Difference Analogue of the Korteweg-de Vries Equation,}  \textit{J. Phys. Soc. Japan}, \textbf{43}  1424 (1977)

\bibitem{Ablowitz} M. J. Ablowitz and J. Ladik, \emph{Nonlinear
differential-difference equations,} \textit{J. Math. Phys.}, \textbf{16} 598 (1975)

\bibitem{date} Date E, Jimbo M and Miwa T, {Method for Generating
Discrete Soliton Equations. I,} \textit{J. Phys. Soc. Japan}, \textbf{51}  4116--4127 (1982).

\bibitem{Hirota:dKP}
R. Hirota,
Discrete Analogue of a Generalized Toda Equation,
\textit{J. Phys. Soc. Jpn.}, \textbf{50} 3785--3791 (1981)

\bibitem{Miwa} T. Miwa,
On Hirota's difference equations,
\textit{Proc. Jpn. Acad. A}, \textbf{58} 9--12 (1982)


\bibitem{suris} Y. B. Suris, \textit{The Problem of Integrable Discretization: Hamiltonian Approach} (Birkh\"auser, Basel, 2003)

\bibitem{bobenko} A.I. Bobenko, J. M. Sullivan, P. Schr\"oder and G.
Ziegler, \textit{Discrete differential geometry}, (Spinger, 2008).

\bibitem{FMO-PJMI} B.-F. Feng, K. Maruno and Y. Ohta, Self-adaptive moving
mesh schemes for short pulse type equations and their Lax pairs, \textit{Pacific Journal of Mathematics for Industry}, \textbf{6} 1--14 (2014).

\bibitem{FMOmultiSP} B.-F. Feng, K. Maruno and Y. Ohta, Integrable
discretization of a multi-component short pulse equation, \textit{J. Math. Phys.} \textbf{56} 043502 (2015).

%\bibitem{fengshort} B.-F. Feng, \emph{Complex short pulse and coupled
%complex short pulse equations,} Physica D, 297:(2015) 62-75.


%\bibitem{fengdis} B. F. Feng, K. Maruno and Y. Ohta, \emph{Self-adaptive
%moving mesh schemes for short pulse type equations and their Lax pairs,}
%Pacific Journal of Mathematics for Industry, 6(1):(2014) 1-14.

\bibitem{Matveev} V. B. Matveev and M. A. Salle, \textit{Darboux
transformations and solitons}, (Springer, Berlin, 1991).

\bibitem{Guo1} B. Guo, L. Ling and Q P. Liu, {Nonlinear Schr\"odinger
equation: generalized Darboux transformation and rogue wave solutions,}
\textit{Phys. Rev. E} \textbf{85} 026607 (2012).

\bibitem{Guo2} B. Guo, L. Ling and Q P. Liu, {High-Order Solutions and
Generalized Darboux Transformations of Derivative Nonlinear Schr\"odinger
Equations,} \textit{Stud. Appl. Math.} \textbf{130} 317-344  (2013).

\bibitem{Taoxu1} M. Li, T. Xu, {Dark and antidark soliton interactions in the nonlocal nonlinear Schr\"odinger equation with the self-induced parity-time-symmetric potentia,} \textit{Phys. Rev. E}, \textbf{91} 033202 (2015).

%M. Li and T. Xu, Phys. Rev. E 91, 033202 (2015)

\bibitem{Taoxu2} T. Xu, S. Lan, M. Li, L.-L. Li, G.-W. Zhang, {Mixed soliton solutions of the defocusing nonlocal nonlinear Schr\"odinger equation,} \textit{Physica D} doi/10.1016/j.physd.2018.11.001.

\bibitem{hassan} H.Wajahat A. Riaz, Mahmood ul Hassan,{Darboux
transformation of a semi-discrete coupled dispersionless integrable system}
, \textit{Commun Nonlinear Sci Numer Simulat} \textbf{48}, 387--397 (2017).

\bibitem{loop-group} C.L. Terng and K. Uhlenbeck, B\"acklund
transformations and loop group actions,\textit{Commun. Pure Appl. Math.} \textbf{53}, 1-75 (2000).

\bibitem{alg-geo} E. Belokolos, A. Bobenko, V. Enol'skij, A. Its and V.B.
Matveev, \textit{Algebro-geometric approach to nonlinear integrable equations}, (Springer, 1994).

\bibitem{Ling} L. Ling, L.C. Zhao and B. Guo, Darboux transformation
and multi-dark soliton for N-component nonlinear Schr\"odinger equations,
\textit{Nonlinearity} \textbf{28}, 3243 (2015).
%\bibitem{alg-geo} E. Belokolos, A. Bobenko, V. Enol\'skij, A. Its, V.B.
%Matveev, \emph{Algebro-Geometric approach to nonlinear integrable equations,}
%Springer, 1994.

\bibitem{BP} D. Bilman and P. D. Miller, A robust inverse scattering transform for the focusing nonlinear Schr\"odinger equation, arXiv:1710.06568, 2017.

\bibitem{BLM}
D. Bilman, L. Ling and P. D. Miller, Extreme Superposition: RogueWaves of Infinite Order and the Painlev?e-III Hierarchy,
arXiv:1806.00545, 2018.

\end{thebibliography}
\end{document}